\documentclass[apjl,letterpaper]{emulateapj}
\usepackage{amsmath,amsfonts,amssymb}
\newcommand{\bit}[1]{\mbox{\textbf{\emph{#1}}}}
\newcommand{\phz}{photo-\emph{z}}
\newcommand{\chgd}[1]{#1}

\begin{document}

\title{Distance, Growth Factor, and Dark Energy Constraints from 
Photometric Baryon Acoustic Oscillation and Weak Lensing
Measurements}
\shorttitle{Distance, Growth Factor, and Dark Energy}
\author{Hu Zhan} 
\author{Lloyd Knox}
\author{J.~Anthony Tyson}
\shortauthors{Zhan, Knox, \& Tyson}
\affil{Department of Physics, University of California, 
Davis, CA 95616; hzhan@ucdavis.edu}

\begin{abstract}
Baryon acoustic oscillations (BAOs) and weak lensing (WL) are 
complementary probes of cosmology.
We explore the distance and growth factor measurements from 
photometric BAO and WL techniques and investigate the roles of the 
distance and growth factor in constraining dark energy.
We find for WL that the growth factor has a great impact on dark 
energy constraints but is much less powerful than the distance.
Dark energy constraints from WL are concentrated in considerably 
fewer distance eigenmodes than those from BAO, with the largest
contributions from modes that are sensitive to the absolute 
distance. Both techniques have 
some well determined distance eigenmodes that are not very sensitive 
to the dark energy equation of state parameters $w_0$ and $w_a$, 
suggesting that they can accommodate additional parameters for
dark energy and for the control of systematic uncertainties.
A joint analysis of BAO and WL is far more powerful than either 
technique alone, and the resulting constraints on the distance and 
growth factor will be useful for distinguishing dark energy and 
modified gravity models. The Large Synoptic Survey Telescope (LSST) 
will yield both WL and angular BAO over a sample of several billion 
galaxies. \chgd{Joint LSST BAO and WL can yield $0.5\%$ level 
precision on ten comoving distances evenly spaced in $\log(1+z)$ 
between redshift $0.3$ and $3$ with cosmic microwave 
background priors from {\it Planck}.
In addition, since the angular diameter distance, which directly
affects the observables, is linked to the comoving distance solely
by the curvature radius in the Friedmann-Robertson-Walker metric
solution, LSST can achieve a pure metric constraint 
of 0.017 on the mean curvature parameter $\Omega_k$ of the universe 
simultaneously with the constraints on the comoving distances.}
\end{abstract}


\section{Introduction}\label{sec:intr}

The luminosity distance--redshift relation measured from type 
Ia supernovae (SNe) has provided a vital piece of evidence for an 
accelerated cosmic expansion in a Friedman-Robertson-Walker 
universe \citep{riess98,perlmutter99a}. Such an acceleration
suggests that either the universe is dominated by a smooth dark
energy component or we need an alternative framework of gravity to 
account for the SN data \citep[e.g.,][]{dvali00,deffayet01}.
Although different answers have vastly different theoretical 
implications, they can only be distinguished indirectly, for 
example, through measurements of the cosmic distance scale as a 
function of redshift and the amplitude of the density fluctuations 
as a function of redshift, i.e., the 
growth factor--redshift relation. One may further derive a 
phenomenological dark energy equation of state (EOS) as a function 
of redshift, $w(z)$, from the distances and growth factors and contrast 
it with predictions from theory.

Besides SNe, baryon acoustic oscillations (BAOs) in the galaxy power 
spectrum \citep*{eisenstein98,cooray01b,blake03,hu03b,linder03b} and 
weak lensing \citep[WL,][]{hu99a,mellier99,bartelmann01,huterer02,
refregier03,song04} can also measure the distance
\citep*{seo03,song05,knox06b,zhan06c}. The sound horizon at the last 
scattering surface determines the BAO scale 
\citep{peebles70,bond84,holtzman89}, which is fixed in comoving space,
so that it can be used as a standard ruler to measure the comoving 
angular diameter distance $D_{A}(z)$ and, with spectroscopy, 
Hubble parameter $H(z)$. 
With precise calibration of the sound horizon from cosmic microwave 
background (CMB) observations, the BAO technique 
is able to measure the absolute distance, whereas the SN 
technique measures the relative distance because of the degeneracy
between the SN intrinsic luminosity and the Hubble constant $H_0$.
The WL technique does not depend on power spectrum features; its 
ability to measure distance originates from its geometric
lensing kernel and the sensitivity of the shear power spectrum 
amplitude to the absolute distance
\citep[e.g.,][hereafter ZK06b]{zhan06e}. 

The WL shear signal 
is a direct indicator of the gravitational potential of all matter, 
luminous or not, so the WL technique has the advantage of being 
able to measure the growth factor of the large-scale structure directly.
For the BAO technique\footnote{Although the BAO features are not 
related to the growth factor, the BAO technique, being a power spectrum 
or correlation function analysis, can potentially measure
the growth factor.}, 
the growth factor is largely degenerate with the 
galaxy clustering bias, though it could be determined from the 
shape of the galaxy power spectrum \emph{if} one knew precisely how 
the amplitude of the linear matter power spectrum affects the slope of 
the nonlinear matter power spectrum \emph{and} how the galaxy bias 
behaves in the mildly nonlinear regime (see Section~\ref{sec:dgcon}). 

The relative importance of the distance and growth factor in constraining
the dark energy EOS with WL has been discussed by \citet{abazajian03}, 
\citet[][hereafter SB05]{simpson05}, \citet*[][hereafter ZHS05]{zhang05},
 and \citet{knox06b}, but their
results are apparently inconsistent: SB05 and 
ZHS05 argue that the distance and growth factor are equally
powerful whereas \citet{abazajian03} and  \citet{knox06b} show
that the growth factor is much less powerful than the distance. 
This inconsistency is found to be caused by the parameter space of 
each analysis: with no other parameters, the 
distance and growth factor place comparable constraints on the dark 
energy EOS parameters, but the growth factor becomes less powerful 
when more parameters are allowed to float (ZK06b).
Because the WL measurements of the distance and growth factor are 
entangled, ambiguity arises on the exact meaning of ``using only the
distance (or growth) information'' without a reconstruction 
of distances and growth factors (actually, their errors). 
In this paper, we give a detailed account of
estimating the distance and growth errors and their roles 
in constraining dark energy.

Throughout this paper, we assume a low-density cold dark matter (CDM)
 universe with the 
following parameters: the dark energy equation-of-state parameters 
$w_0$ and $w_a$ as defined by $w(z) = w_0 + w_az(1+z)^{-1}$
\citep{chevallier01}, the 
matter density $\omega_{m}$, the baryon density $\omega_{b}$, the 
angular size of the sound horizon at the last scattering surface 
$\theta_{s}$, the equivalent matter fraction of curvature 
$\Omega_k$, the optical depth to scattering by electrons in the 
reionized inter-galactic medium, $\tau$, the primordial helium mass 
fraction $Y_{p}$, the spectral index $n_{s}$ and running 
$\alpha_{s}$ of the primordial 
scalar perturbation power spectrum, and the normalization of the 
primordial curvature power spectrum $\Delta_R^2$ at 
$k = 0.05\,\mbox{Mpc}^{-1}$. For reasons stated in 
Section~\ref{sec:dgpar}, only a subset of these parameters are
retained when distance and growth parameters are present. 
The fiducial model is taken from 
the 3-year {\it WMAP} data \citep{spergel07}: 
($w_0$, $w_a$, $\omega_{m}$, $\omega_{b}$, $\theta_{s}$,
$\Omega_k$, $\tau$, $Y_{p}$, $n_{s}$, $\alpha_{s}$,
$\Delta_R^2$) = 
(-1, 0, 0.127, 0.0223, 0.596$^\circ$, 0, 0.09, 0.24, 0.951, 0,
$2.0 \times 10^{-9}$). The reduced Hubble constant $h = 0.73$ and 
the present equivalent matter fraction of dark energy 
$\Omega_{X} = 0.76$ are implicit in this parametrization, 
meaning that either one of them can replace $\theta_{s}$ or
any parameter that affects $\theta_{s}$.

The rest of the paper is organized as follows. Section~\ref{sec:wb}
provides a brief introduction to the power spectrum analysis and 
error forecast with BAO and WL. For the investigation that follows,
we assume a photometric redshift (\phz{}) survey based on the Large 
Synoptic Survey Telescope\footnote{See \url{http://www.lsst.org}.} 
\citep[LSST,][]{ivezic08} and consider only \phz{} BAO and WL 
tomography. We compare BAO and WL constraints on the distance and 
growth factor in Section~\ref{sec:dgch} and examine the roles of 
distance and growth for studying dark energy in Section~\ref{sec:pde}.
The conclusions are drawn in Section~\ref{sec:con}. 
Since our main goal is to understand the intrinsic properties of the 
distance and growth factor measured from the two techniques and their 
applications in dark energy studies, we neglect systematics 
in Sections~\ref{sec:dgch} and \ref{sec:pde}. To be more 
realistic, we include in Section~\ref{sec:con} a brief discussion of 
the constraints on the distance and growth factor from a joint 
analysis of LSST BAO and WL with conservative estimates of 
systematic uncertainties in \phz{}s 
and power spectra. For more detailed accounts of various
systematics, see \citet*{huterer05a,white05,ma06,huterer06,jain06,
zhan06d,eisenstein07b,guzik07,zentner08}.

\section{Weak Lensing and Baryon Acoustic Oscillations} \label{sec:wb}

The projected lensing potential $\phi(\boldsymbol{\theta})$ transforms
a small displacement $\Delta \boldsymbol{\beta}$ in the source plane 
into a small displacement $\Delta\boldsymbol{\theta}$ 
in the image plane via \citep{kaiser92,bartelmann01}
\[
\Delta \boldsymbol{\beta} =
\left(\begin{array}{cc} 1-\kappa-\gamma_1 & -\gamma_2 \\
-\gamma_2 & 1-\kappa+\gamma_1\end{array}\right) 
\Delta \boldsymbol{\theta}, 
\]
where the convergence $\kappa=\nabla^2\phi/2$, the shear components
$\gamma_1=(\partial^2\phi/\partial\theta_1^2-
\partial^2\phi/\partial\theta_2^2)/2$ and 
$\gamma_2=\partial^2\phi/\partial\theta_1\partial\theta_2$. The
shear $\gamma=\gamma_1+i\gamma_2$ represents the distortion of an image
without magnification, so that it can be inferred from the average 
ellipticity of galaxies within an appropriate window under the 
assumption that galaxies are randomly oriented in the absence of 
lensing. One may decompose a shear map into E-modes and B-modes, but,
to first order, gravitational lensing from density fluctuations does 
not induce B-modes. Therefore, we only consider the E-mode shear 
statistics here.

The angular power spectra of the shear 
$\gamma(\boldsymbol{\theta})$ and galaxy number density 
$n(\boldsymbol{\theta})$  can be written as 
\citep{hu04b,zhan06d}
\begin{equation} \label{eq:aps} 
P_{ij}^{XY}(\ell) = \frac{2\pi^2}{c\ell^3} \int_0^\infty d z\, H(z) 
D_{A}(z) W_i^X(z) W_j^Y(z) \Delta^2_\delta(k; z),
\end{equation}
where lower case subscripts correspond to the tomographic bins, upper 
case superscript labels the observables, e.g., $X= \mbox{g}$ for 
galaxies or $\gamma$ for shear, 
$\Delta^2_\delta(k;z)$ is the dimensionless power spectrum
of the density field, \chgd{$D_A(z)$ is the comoving angular diameter 
distance of redshift $z$ as viewed from redshift 0 
[the same as the comoving distance, 
$D(z)$, between redshift 0 and $z$ in a flat universe],} and 
$k = \ell/D_{A}(z)$. In the linear regime, the power spectrum is 
scaled by the growth factor $G(z)$
\[
\Delta^2_\delta(k;z) = G^2(z)\Delta^2_\delta(k;0) /G^2(0).
\]
The window functions are
\begin{eqnarray} 
W_i^{g}(z) &=& b(z)\frac{n_i(z)}{\bar{n}_i}, \nonumber \\
W_i^\gamma(z) &=& \frac{3}{2}
\frac{\Omega_{m}H_0^2}{H(z)}\frac{D_{A}(z)}{a\,c} 
\int_z^\infty \! d z'\, \frac{n_i(z')}{\bar{n}_i}
\frac{D_{A}(z,z')}{D_{A}(z')}, \nonumber
\end{eqnarray}
where $b(z)$ is the linear galaxy clustering bias,  
$\Omega_{m}=\omega_{m}/h^2$, \chgd{and $D_{A}(z,z')$ is the comoving 
angular diameter distance of redshift $z'$ as viewed from $z$
[the same as the comoving distance, $D(z, z')$, between
redshift $z$ and $z'$ in a flat universe].}
The galaxy redshift distribution $n_i(z)$ in the
$i$th tomographic bin is an average of the underlying 
three-dimensional galaxy distribution over angles, and the mean
surface density $\bar{n}_i$ is the total number of galaxies per
steradian in bin $i$. The binning for WL need not be the same as
that for BAO. 
The irreducible occurrence of the Hubble parameter $H(z)$ in 
equation (\ref{eq:aps}) for BAO is due to the fact that the galaxy
number density $n_i(z)$ is measured in redshift not in distance.

Equation (\ref{eq:aps}) is applicable to the
shear power spectrum for WL ($X=Y=\gamma$), galaxy power spectrum for
BAO ($X=Y=\mathrm{g}$), and galaxy-shear power spectrum ($X=\mathrm{g}$ 
and $Y=\gamma$). Although we focus on gaining a 
better understanding of the BAO and WL techniques by exploring their 
differences, we must emphasize that a joint analysis of BAO and WL 
with all the three types of power spectra \citep{hu04b,zhan06d}
is far more powerful than either technique alone. 

We parametrize the underlying galaxy redshift distribution as
\citep{wittman00}
\[
n(z) \propto z^\alpha \exp\left[-(z/z^*)^\beta\right]
\]
and adopt the values $\alpha = 2$, $z^*=0.5$, and $\beta=1$ with a 
projected galaxy number density of $n_{\rm tot} = 50$ per square 
arcminute for LSST. 
The galaxy distribution $n_i(z)$ in the $i$th bin 
is sampled from $n(z)$ by \citep{ma06,zhan06d}
\[
n_i(z) = n(z) \mathcal{P}(z_{{p},i}^{\rm B}, z_{{p},i}^{\rm E}; z),
\]
where the subscript p denotes \phz{} space, 
$z_{{p},i}^{\rm B}$ and $z_{{p},i}^{\rm E}$ define the 
extent of bin $i$, and $\mathcal{P}(a,b;z)$ is the probability of 
assigning a galaxy that is at true redshift $z$ to 
the \phz{} bin between $z_{p} = a$ and $b$.
We approximate the \phz{} error to be Gaussian with bias $\delta z$ 
and rms $\sigma_z=\sigma_{z0} (1+z)$, and the probability becomes
\begin{eqnarray} \nonumber
\mathcal{P}(z_{{p},i}^{\rm B}, z_{{p},i}^{\rm E}; z)
&=& I(z_{{p},i}^{\rm B}, z_{{p},i}^{\rm E}; z)
 / I(0,\infty; z), \\ \nonumber
I(a, b; z) &=& \frac{1}{\sqrt{2\pi}\,\sigma_z} \int_a^b d z_{p} 
\,\exp\left[-\frac{(z_{p} - z - \delta z)^2}{2\sigma_z^2}\right].
\end{eqnarray} 
The normalization $I(0,\infty; z)$ implies that galaxies with 
a negative \phz{} have been excluded from $n(z)$. Although 
uncertainties in the
\phz{} parameters $\delta z$ and $\sigma_z$ have a large impact on 
the dark energy constraints from WL \citep{huterer06,ma06,zhan06d}, 
such systematic effects are not relevant to how WL 
probes dark energy. Hence, we fix $\delta z = 0$ and 
$\sigma_{z} = 0.05(1+z)$ per galaxy. In Section~\ref{sec:con} we 
show results with \phz{} and power spectrum systematics.

Assuming that the observables, e.g., the shear map 
$\gamma(\boldsymbol{\theta})$ in multipole space, and the likelihood
of the parameters of interest, $\bit{q}$, near the fiducial 
model are both Gaussian, one can use the Fisher information matrix 
to estimate the errors of the parameters from the covariance, 
$\bit{C}$, of the observables. 
In summary, the Fisher matrix is given by \citep{tegmark97b}
\begin{equation} \label{eq:trfish}
F_{\alpha\beta} = f_{\rm sky} \sum_\ell \frac{2\ell + 1}{2} {\rm Tr} 
\bit{C}_\ell^{-1} \frac{\partial \bit{C}_\ell}{\partial q_\alpha} 
\bit{C}_\ell^{-1} \frac{\partial \bit{C}_\ell}{\partial q_\beta}.
\end{equation}
For BAO and WL tomography, we have
\[
(\bit{C}_\ell^{XY})_{ij} = P_{ij}^{XY}(\ell) + \delta_{XY}^{\rm K}
\delta_{ij}^{\rm K} X_{\rm rms}^2 \bar{n}_i^{-1},
\]
where $\delta_{ij}^{\rm K}$ is the Kronecker delta function, 
$\gamma_{\rm rms}\sim 0.2$ is the rms shear per galaxy for WL, and
${\rm g_{rms}}\equiv 1$ for BAO. 
The Gaussian likelihood of the parameters is
\begin{equation} \label{eq:probq} \nonumber
\mathcal{L}(\bit{q}) \propto \left|\bit{F}^{-1}\right|^{-1/2} 
\exp\left[-\frac{1}{2}\left(\bit{q}-\bit{q}_{f}\right)^{\rm T} 
\bit{F} \left(\bit{q} - \bit{q}_{f}\right)\right], 
\end{equation}
where $\bit{q}_{f}$ corresponds to the fiducial model and
$\bit{F}^{-1}$ is the covariance of the parameters. 
The minimum marginalized $1\sigma$ error of $q_\alpha$ is 
$\sigma(q_\alpha) = (\bit{F}^{-1})_{\alpha\alpha}^{1/2}$. 
Independent Fisher matrices are 
additive; a prior on $q_\alpha$, $\sigma_{P}(q_\alpha)$, 
can be introduced via $F_{\alpha\alpha}^{\prime} = F_{\alpha\alpha}
+\sigma_{P}^{-2}(q_\alpha)$. 
A Fisher matrix of the parameters
$\bit{q}$ can be projected onto a new set of parameters 
$\bit{p}$ via
\begin{equation} \label{eq:projfsh}
F_{\mu\nu}^{\prime}= \sum_{\alpha,\beta}
\frac{\partial q_\alpha}{\partial{p_\mu}} F_{\alpha\beta}
\frac{\partial q_\beta}{\partial{p_\nu}}.
\end{equation}

\section{Distance and Growth Factor Constraints} \label{sec:dgch}

The distance--redshift and growth factor--redshift relations are the two
most important diagnostics of the driving mechanism of the accelerated
cosmic expansion, be it dark energy or modified gravity. 
Various dark energy probes are essentially probes of these relations
(and hence they are modified gravity probes as well). 
Most models of dark energy and modified gravity do not modify the form 
of equation (\ref{eq:aps}). As such, reconstructing the 
distance and growth factor 
from the shear and galaxy power spectra is fairly model-independent,
which is useful for distinguishing those models (\citealt*{lue04}; 
\citealt{song05}; \citealt*{ishak06}; \citealt{knox06b}; 
\citealt{jain07}; but cf. \citealt{linder04}; \citealt{bertschinger08}). 

To reconstruct the distance and growth factor, one would set a series
of distance and growth parameters covering a range of redshifts, 
interpolate the continuous distance and growth functions from a 
realization of these parameters, assign a probability of the observed
data given the particular set of distance, growth, and other parameters,
and repeat the process for enough realizations to map the posterior 
distribution of all the parameters. This is rather involved. However,
if the purpose is to roughly estimate the errors without the actual
observational data, one often approximates the posterior distribution 
of the parameters as a multivariate Gaussian around its peak and 
applies the Fisher matrix analysis to infer the errors.
We show here an example of estimating the errors of 
reconstructed distances and growth factors from BAO and WL, both 
separately and in combination.

\subsection{Parameters} \label{sec:dgpar}
There is a minor ambiguity in the growth factor of the large-scale
structure, and we choose 
the convention that $G(z) = (1+z)^{-1}$ in an Einstein-de Sitter 
universe but $G(0)\ne 1$ in general. For distance parameters, 
it is convenient to use the comoving 
distance $D(z)$ instead of the comoving angular diameter distance 
$D_{A}(z)$. 
The reason is that while $D(z,z') = D(z') - D(z)$ holds all the time,
$D_{A}(z,z')\ne D_{A}(z') - D_{A}(z)$ 
in the presence of curvature:
\begin{equation}  \label{eq:DA-Ok}
D_{A}(z,z') = \left\{ \begin{array}{ll}
K^{-1/2} \sin[D(z, z') K^{1/2}] & \quad K > 0 \\
D(z, z') & \quad K = 0 \\
|K|^{-1/2} \sinh[D(z, z') |K|^{1/2}] & \quad K < 0 \end{array} \right.,
\end{equation}
where $K=-\Omega_k(H_0/c)^2$. Separately, the freedom of 
the Hubble parameter, $H(z) = c (dD/dz)^{-1}$, in the galaxy power 
spectra of equation (\ref{eq:aps}) could be unphysically restricted by 
the interpolation scheme. A conceptually simple remedy is 
to replace distance parameters $D_i$ with Hubble parameters $H_i$ and 
then project $H_i$ constraints onto $H_0$ and $D_i$ constraints. 

We assign 15 Hubble parameters $H_i$ and 15 growth parameters 
$G_i$ ($i = 0 \ldots 14$) at redshifts evenly spaced in 
$\log(1+z)$ from $z_0 = 0$ to $z_{14}=5$. 
The Hubble parameter $H(z)$ and growth factor $G(z)$ are
then spline-interpolated from $H_i$s and $G_i$s. The comoving distance
parameters $D_i$ are assigned at the same redshifts except that 
$D_0 \equiv 0$ is replaced by $H_0$. Note that our distance 
parameters are in units of Mpc without $h$. For brevity, 
we only present the projected results without referring
to $H_i$ and the intermediate step.

Given a cosmological model for the distance and growth factor, 
the parameter $\theta_{s}$ is a function of other explicit and implicit
parameters that affect the distance and linear size of the sound 
horizon at the last scattering surface. This means that, in the 
absence of the distance and growth parameters, the numerical
derivatives in equation (\ref{eq:trfish}) with respect to each 
parameter must be 
taken with the implicit parameters $h$ and $\Omega_X$ perturbed
in unison to keep other explicit parameters fixed. A consequence is
that even though the physical matter density $\omega_m$, baryon
density $\omega_b$, and $\theta_s$ do not directly affect the 
distance and growth factor, they do so indirectly through the 
implicit parameters.
As one incorporates the distance and growth parameters, the physical
model that links the cosmological parameters to the distance and 
growth factor is no longer in effect, so information about 
the distance and growth factor must be carefully removed from the 
cosmological parameters. 

Parameters that affect the distance or growth factor without altering 
the normalization or the shape of the matter power spectrum may not 
be included in the forecast. Such parameters include the dark energy 
EOS parameters and $\theta_s$. The curvature parameter $\Omega_k$ has
a negligible effect on the slope of the matter power spectrum. It
affects the angular diameter distance through its impact on the
comoving distance, which is replaced by $D_i$s, and via 
equation (\ref{eq:DA-Ok}), which is a property of the 
Friedmann-Robertson-Walker (FRW) metric.
As such, we keep $\Omega_k$ for (and only for) calculating $D_{A}$ 
from $D_i$s using equation (\ref{eq:DA-Ok}). Since the curvature 
parameter here has no role in the matter power spectrum, $D_i$, or 
$G_i$, we label it as $\Omega_k^\prime$ to distinguish from the 
real curvature parameter. 

\chgd{Our approach is similar to that in
\citet{bernstein06} where $D_A(z,z')$ is expressed in terms of 
$D_A(z)$, $D_A(z')$, and, to first order, $\Omega_k'$. The constraint 
on $\Omega_k^\prime$ will hold for any model that preserves the FRW 
metric and (the form of) equation (\ref{eq:aps}), whereas curvature 
constraints from exploiting the full functional dependence of the 
angular diameter distance or luminosity distance on $\Omega_k$ 
\citep[e.g.,][]{spergel07} are valid only for a particular 
cosmological model. For this reason, measurements of 
$\Omega_k^\prime$ are considered pure metric tests for curvature 
\citep{bernstein06}.}

The rest of the cosmological parameters do not affect the distance 
or growth factor. However, if the growth factor is already accounted 
for by the code that calculates the matter power spectrum or transfer 
function \citep[e.g., {\sc cmbfast},][]{zaldarriaga00}, one must
use proper combinations of the code input parameters to obtain the
derivatives for the Fisher matrix without altering the growth factor.

In summary, our parameter set for forecasting distance and growth 
errors includes $\omega_{m}$, $\omega_{b}$, $\Omega_k^\prime$, 
$Y_{p}$, $\tau$, $n_{s}$, $\alpha_{s}$, $\Delta_R^2$, 
$H_0$, $D_i$ ($i = 1\ldots 14$), and $G_i$ 
($i = 0\ldots 14$). For BAO, additional 15 linear 
galaxy bias parameters $b_i$ are included, which are assigned
at the same redshifts as the growth parameters. 
In actual calculations, we use $\ln H_0$, $\ln D_i$, $\ln G_i$, and
$\ln b_i$ as parameters, so their constraints are reported in 
fractional form.

\chgd{We assume 
a fiducial model for the linear galaxy bias $b(z) = 1+0.84z$, which
is estimated from the simulation results in \citet{weinberg04}.
The exact value of $b(z)$ is not important for our purpose, though
a higher bias does produce stronger signals (galaxy power spectra) 
and hence tighter parameter constraints. The dependence of the dark 
energy EOS error on the power spectrum amplitude can be found in 
\citet{zhan06d}. The nonlinear matter power spectrum is calculated
using the \citet{peacock96} fitting formula. A direct application 
of the fitting formula to the CDM power spectrum would cause a large
shift of the BAO features. In addition, it has difficulty processing 
power spectra that have an oscillating logarithmic slope. Thus, we
first calculate a linear matter power spectrum with no BAO features
that otherwise matches the CDM power spectrum \citep{eisenstein99a},
then take the ratio between the nonlinear and linear matter power 
spectra with no BAO features, and finally apply this ratio
to the linear CDM power spectrum to obtain the nonlinear CDM power
spectrum \citep[see also][]{eisenstein05,zhan06d}.
}

We modify {\sc cswab}\footnote{Available at 
\url{http://software.hzhan.net}.} \citep{zhan06d,zhan08a} to calculate 
the Fisher matrices for 10-bin ($0 \le z_{p} \le 3.5$) WL and 
20-bin ($0.15 \le z_{p} \le 3.5$) BAO tomographic measurements
over 20,000 square degrees.
To reduce the impact of nonlinear evolution as well as baryonic 
effects on the small-scale matter power spectrum 
\citep{white04,zhan04c,hagan05,jing06,rudd08}, we limit the maximum 
multipole $\ell_{\rm max}$ to 2000 for WL and 3000 for BAO [the 
latter must also satisfy the condition 
$\Delta_\delta^2(\ell/D_{A};z) < 0.4$]. 
At the low multipole end, we exclude modes with $\ell < 40$, which 
could be affected by dark energy clustering \citep{hu04c,song04}. Since 
$D_i$, $G_i$, and $b_i$ are not well constrained by the data at 
the low- and high-redshift ends, we impose extremely weak priors of 
$\sigma_{P}(\ln D_i)=\sigma_{P}(\ln G_i)=\sigma_{P}(\ln b_i)
=100$, i.e., 10,000\% error, just to avoid vanishing diagonal elements 
in the Fisher matrices.

\begin{figure*}
\centering
\epsscale{0.86}
\plotone{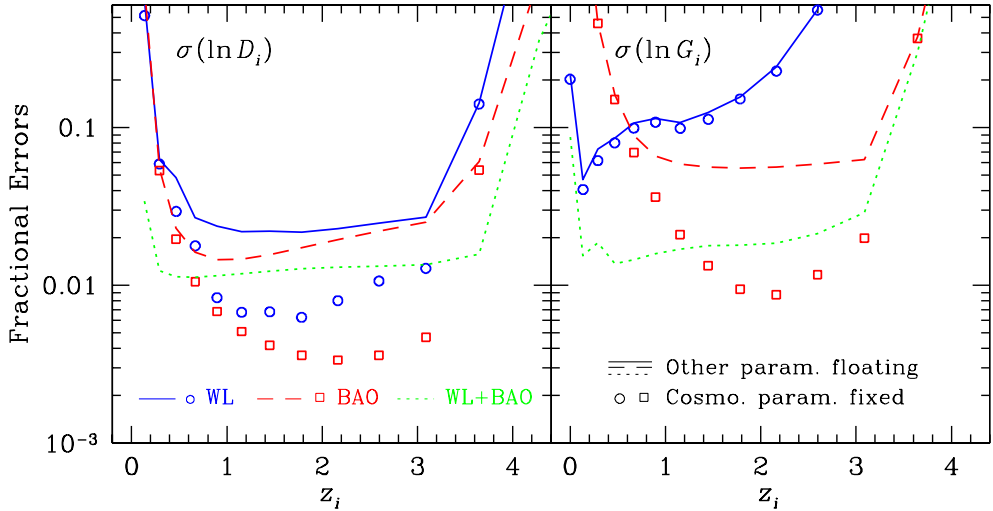}
\caption{Marginalized $1\sigma$ errors on the comoving distance 
(left panel) and growth factor 
(right panel) parameters from LSST WL (solid line 
and open circles), BAO (dashed line and open squares), and joint WL 
and BAO (dotted line). \chgd{The maximum multiple used for WL is 
2000, and that for BAO is 3000 [with the additional requirement
$\Delta_\delta^2(\ell/D_{A};z) < 0.4$].}
The growth parameters, $G_0 \ldots D_{14}$, are evenly spaced in 
$\log(1+z)$ between $z = 0$ and 5, and the distance parameters, 
$D_1 \ldots D_{14}$, start at $z_1 = 0.14$ (see text for details).
\chgd{The error of each distance (growth) parameter is marginalized 
over all the other parameters including growth (distance) parameters
and other distance (growth) parameters.} 
Lines represent results with cosmological parameters 
floating, i.e., marginalized with weak priors, and symbols are for 
results with the cosmological parameters fixed.
\label{fig:dgcon}}
\end{figure*}

\subsection{Distance and Growth Factor Constraints} \label{sec:dgcon}

How well can LSST measure the distance and growth factor given its 
statistical errors?
Fig.~\ref{fig:dgcon} presents the estimated errors on the distance 
(left panel) and growth (right panel) parameters from LSST WL (solid 
lines and open circles), BAO (dashed lines and open squares), and 
joint BAO and WL (dotted lines). 
\chgd{The error of each distance (growth) parameter is marginalized 
over all the other parameters including the growth (distance) 
parameters and other distance (growth) parameters.} 
The lines show the results with
weak independent Gaussian priors on cosmological parameters:
$\sigma_{P}(\ln\omega_{m})=\sigma_{P}(\ln\omega_{b})=
\sigma_{P}(\Omega_k^\prime) = 
\sigma_{P}(n_{s})=\sigma_{P}(\alpha_{s}) = 0.05$,
$\sigma_{P}(\tau)=0.01$, $\sigma_{P}(Y_{p})=0.02$,
and $\sigma_{P}(\ln\Delta_R^2)=0.1$. These priors are rather
weak compared to what can be achieved with {\it Planck}, except that 
$\Omega_k^\prime$ is not constrained by CMB unless a 
particular cosmological model and appropriate priors are assumed. 
In addition,
they do not introduce extra correlations between the parameters. 
The errors on the galaxy bias parameters (not shown) trace the errors 
of BAO growth parameters in this case.
The symbols assume that the cosmological parameters 
are known precisely. A discussion on the sensitivity of these distance 
and growth factor constraints to individual cosmological parameters is
given in Appendix~\ref{sec:dcldq}.

It is not surprising that BAO places tighter constraints 
on the distance parameters than WL  
\citep[see also][for results of either technique]{seo03,song05,knox06b,
zhan06c}, but one may not expect that the former determines
the growth parameters better than the latter at $z \gtrsim 0.7$.
It has been demonstrated that \phz{} BAO measurements (without 
cross-bin power spectra) can constrain the combination of
the galaxy bias and growth factor $b_iG_i$ to several percent
\citep{zhan06c}, but the degeneracy between $b_i$ and $G_i$ could 
have left them undetermined individually. 
However, $b_i$ and $G_i$ are not completely degenerate here; 
the linear galaxy bias is treated merely as a multiplicative 
factor in the galaxy power spectrum, whereas the growth factor (among 
others) determines both the amplitude and the shape of the nonlinear 
matter power spectrum \citep{peacock96}. 
Since the BAO technique is sensitive to the shape
of the power spectrum, it is able to partially break the 
degeneracy and constrain both $G_i$ and $b_i$. 
\chgd{This also means that the BAO constraints on 
the growth factor can be sensitive to the number of nonlinear modes 
included. For example, with $\ell_{\rm max} = 1000$ and all the other 
parameters floating, the BAO error on the growth factor at $z = 3.1$ 
in the right panel of Fig.~\ref{fig:dgcon} increases from 
$0.064$ to $0.21$, though that at $z=1.2$ increases by only 
15\%. The impact of the nonlinear information is compounded by the 
fact that including more modes, regardless whether they are linear 
or nonlinear, always improves the constraints on non-degenerate 
parameters. If the trace term in equation (\ref{eq:trfish}) is 
roughly scale-independent, the Fisher matrix will scale as 
$\ell_{\rm max}^2$. This alone could claim a factor of 3 increase
in the error as $\ell_{\rm max}$ decreases from 3000 to 1000.
The small difference of the errors at low redshift is due to the fact 
that the highest multipole there is already limited by the condition
$\Delta_\delta^2(\ell/D_{A};z) < 0.4$, a measure to 
exclude fully nonlinear modes. On the other hand, if we relax the 
conditions to $\Delta_\delta^2(\ell/D_{A};z) < 1$ and 
$\ell_{\rm max} = 6000$, the errors on the growth factor reduce to 
$\sim 0.04$ between $z = 0.29$ and $3.1$.}

\chgd{It may seem quite optimistic to rely on the nonlinear power 
spectrum, even if restricted to mildly nonlinear regime, for 
determining the growth factor.} However, this is not a serious 
concern, because the growth factor contributes 
little to BAO constraints on dark energy (see Section~\ref{sec:dgde}).
Moreover, it has been demonstrated that there is
no significant difference between the WL dark energy constraints 
with a scale-free matter power spectrum or a nonlinear CDM power
spectrum if the shot noise is negligible and that nonlinear
evolution helps WL only because it boosts the signal on small
scales where the statistical uncertainty is limited by the shot 
noise (ZK06b). Nevertheless, Fig.~\ref{fig:dgcon} suggests that
the galaxy power spectra could potentially constrain dark energy and 
alternative gravity models with both distance and growth factor 
measurements as WL does \emph{if} one understood the galaxy bias 
and nonlinear evolution well through numerical simulations
\citep[e.g.,][]{cen00,weinberg04,heitmann05}, semi-analytic modeling 
\citep[e.g.,][]{ma00,peacock00,croton07}, perturbative calculations 
\citep[e.g.,][]{jain94,jeong06}, and direct measurements from 
higher-order statistics of the galaxy distribution 
\citep*{fry94,mo97,verde02,gaztanaga05,sefusatti07}.

The joint BAO and WL analysis reduces the errors on the distance and
growth factor parameters. Although the improvement on the marginalized 
distance errors is moderate, the resulting constraints on the dark 
energy EOS parameters are much tighter than those from either 
technique alone (see Section~\ref{sec:deigen}). 
The marked improvement on the growth factor
parameters is due to the fact that the shear power spectra, 
galaxy--shear power spectra, and galaxy power spectra have different 
dependencies on the galaxy bias ($b^0$, $b$, and $b^2$, respectively). 
This further lifts the degeneracy between $G_i$ and $b_i$, 
allowing significantly better determinations of $G_i$.

\subsection{Metric Test of Curvature}

The mean curvature of the universe is of great theoretical interest.
It is estimated that future large-scale BAO, SN, and WL surveys can 
constrain $\Omega_k$ (or $\Omega_k h^2$) to better than $10^{-3}$ 
with the assumption of matter dominance at $z \gtrsim 2$ and 
precise independent distance measurements
at $z\gtrsim 2$ and at recombination \citep{knox06a} or with a specific 
dark energy EOS $w(z) = w_0 + w_az(1+z)^{-1}$ \citep{knox06c,zhan06d}. 
However, if one assumes only the Robertson-Walker metric without 
invoking the dependence of the comoving distance on cosmology, then 
the pure metric constraint on curvature from a simple combination of
BAO and WL becomes $\sigma(\Omega_k^\prime) \simeq 0.04
f_{\rm sky}^{-1/2}(\sigma_{z0}/0.04)^{1/2}$ \citep{bernstein06}.

Our result for $\Omega_k^\prime$ from LSST WL or BAO alone is not 
meaningful, but the joint analysis of the two leads to
$\sigma(\Omega_k^\prime)=0.015$ with the aforementioned weak priors
except that no prior is applied to $\Omega_k^\prime$. 
The error on $\Omega_k^\prime$ changes only mildly with the number 
of parameters used, e.g., $\sigma(\Omega_k^\prime)=0.017$ for 100 
$G_i$s and 99 $D_i$s, even though errors of (the eigenmodes of)
the distance and growth factor 
scale roughly as the square root of the number of parameters. This 
is expected, because the mapping between the underlying comoving 
distance and the observable angular diameter distance through
$\Omega_k^\prime$ does not depend on the number of parameters.

\emph{Planck} will provide an accurate measurement of
the angular diameter distance to the last scattering surface, and 
it can reduce the error moderately from $\sigma(\Omega_k^\prime)=0.015$
to $0.013$. Separately, the constraint weakens to 
$\sigma(\Omega_k^\prime)=0.017$ with conservative estimates of 
systematics in \phz{}s and power spectra expected for LSST 
(see Section~\ref{sec:con} for more details).
This is better than the forecast derived from the shear power spectra 
and galaxy power spectra in \citet{bernstein06}, because
we include in our analysis more information: the
galaxy--shear power spectra.

\section{Probing Dark Energy with Distance and Growth Factor} 
\label{sec:pde}

With the distance and growth factor constraints from the last section, 
one can now answer the questions about the roles of the distance and 
growth factor in probing dark energy and the difference between the 
distance measurements from BAO and WL.

To assess the utility of distance and growth factor in constraining 
dark energy, we adopt a generic parametrization of the dark energy 
EOS $w(z) = w_0 + w_a z (1+z)^{-1}$ and project the distance and
growth factor errors onto $w_0$ and $w_{a}$.
We use the product of $1\sigma$ errors, 
$\sigma(w_{p})\times\sigma(w_a)$, to gauge dark energy constraints, 
where the error $\sigma(w_{p})$ is equal to that on $w_0$ with $w_a$ 
being held fixed \citep{huterer01,hu04b}. This error product (EP) is 
proportional to the area of the error ellipse in the $w_0$--$w_a$ plane. 
The figure of merit in the
Dark Energy Task Force report \citep{albrecht06} is the reciprocal 
of the product of $2\sigma$ errors on $w_p$ and $w_a$.

\subsection{Roles of Distance and Growth Factor}
 \label{sec:dgde}

The relative power of the distance and growth factor in 
constraining the dark energy EOS has been studied in a number of ways. 
For example, SB05 and ZHS05 split the dark
energy EOS into two constant components: $w_d$ that affects only the 
distance and $w_g$ that affects only the growth factor.  They find that
the constraints $\sigma(w_d)$ from the distance and $\sigma(w_g)$ 
from the growth factor are comparable. However, \citet{knox06b}, by 
projecting estimated distance and growth factor errors onto a constant 
EOS $w$, mention that the constraint on $w$ is due mostly to 
the distance. ZK06b explores this subject in yet another way; they 
compare the EPs with cosmological information 
removed from either the distance or the growth factor. 
Their approach is equivalent 
to fixing $D_i$ ($G_i$) and then projecting the errors of $G_i$ ($D_i$)
onto cosmological parameter space, though they bypass the intermediate
stage of estimating $D_i$ and $G_i$ errors. 
ZK06b shows that WL distances are 
much more powerful than WL growth factors in constraining $w_0$ and $w_a$
if other cosmological parameters are marginalized with weak priors 
and that they are comparable if the other cosmological parameters are
fixed. 

Because of the different methods employed, it is not 
straightforward to synthesize the above mentioned results. 
Our multi-staged approach is compatible with that in \citet{knox06b},
and we obtain fully consistent results with those in ZK06b when their
method is applied to the $D_i$ and $G_i$ errors in Section 
\ref{sec:dgcon}. Hence, we only need to explore the results for
the method in SB05 and ZHS05.

\begin{figure}
\centering
\epsscale{1}
\plotone{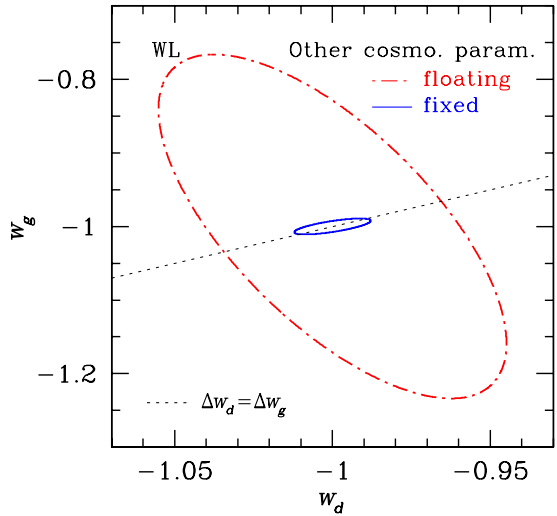}
\caption{WL $1\sigma$ error contours of the dark energy EOS parameters 
$w_d$ and $w_g$, which artificially split the EOS into
a component that affects the distance only ($w_d$) 
and another that affects the growth factor only ($w_g$). 
If all the other cosmological parameters are allowed to float with 
weak priors (dot-dashed line), the constraint on $w_d$ from $D_i$ 
will be much stronger than that on $w_g$ from $G_i$, whereas if all the 
other cosmological parameters are fixed (solid line), 
the constraints on $w_d$ and $w_g$ will become comparable. 
\label{fig:wdwg}}
\end{figure}

We project the WL Fisher matrix in Section \ref{sec:dgcon} onto 
($w_d$, $w_g$, $\omega_{m}$, $\omega_{b}$, $\theta_{s}$,
$\Omega_k$, $\tau$, $Y_{p}$, $n_{s}$, $\alpha_{s}$,
$\Delta_R^2$) using equation (\ref{eq:projfsh}) with 
$d D_i/d w_g = d G_i/d w_d = 0$.
The resulting $1\sigma$ error contours of $w_d$ and $w_g$ are 
shown in Fig.~\ref{fig:wdwg} with
$\sigma(w_d) = 0.036$ and $\sigma(w_g) = 0.15$ when all the other 
cosmological parameters are marginalized with weak priors (dot-dashed 
line) and $\sigma(w_d) = 0.0080$ and $\sigma(w_g) = 0.0071$ when all
the other cosmological parameters are fixed (solid line). 
The latter result is roughly consistent with 
$\sigma(w_d)/\sigma(w_g)\simeq 1/2$ from a single-bin shear power 
spectrum in SB05 and 
$\sigma(w_d)/\sigma(w_g)\simeq 1$ in ZHS05, though the 
orientation of the error contour in ZHS05 is nearly 
orthogonal to that in Fig.~\ref{fig:wdwg} and that in SB05.
\chgd{To examine if the differences of the assumed surveys could 
explain this discrepancy, we re-calculate the constraints with the 
survey parameters used in SB05 ($f_{\rm sky} = 0.024$, 
$\alpha=\beta=2$, $z^*=1.13$, $n_{\rm tot}= 100\,\mbox{arcmin}^{-2}$, 
$\ell_{\rm max} = 10^4$, no tomography, \& no redshift error). 
The resulting errors, $\sigma(w_d)=0.027$ and $\sigma(w_g)=0.032$ 
with all the other parameters fixed, and the orientation of the 
error contour are in good agreement with those in SB05 
except that our $\sigma(w_g)$ is $\sim 40\%$ smaller. 
We are not able to reproduce Fig.~3 of ZHS05 
($f_{\rm sky} = 0.1$, $\alpha=2$, $\beta=1.5$, $z^*=1$, 
$n_\mathrm{tot}= 100\,\mbox{arcmin}^{-2}$, $\ell_{\rm max} = 200$, 
20-bin tomography, \& $\sigma_{z0}=0.05$); our errors on $w_d$ and
$w_g$ remain positively correlated with comparable size 
[$\sigma(w_d)=0.051$ \& $\sigma(w_g)=0.081$] when all the 
other parameters are fixed and anti-correlated with much smaller 
error on $w_d$ [$\sigma(w_d)=0.25$ \& $\sigma(w_g)=1.2$]
when all the other parameters are marginalized with weak priors. 
We note that even though \emph{Planck} will place very tight 
constraints on CMB-related parameters, it does not have the same 
effect as fixing those parameters, e.g., we obtain
$\sigma(w_d) = 0.022$ and $\sigma(w_g) = 0.099$ for LSST with
\emph{Planck} priors. Therefore, 
the distance should be at least several times more powerful than the 
growth factor in constraining the dark energy EOS for WL tomography. }

\begin{deluxetable}{l l l c c c c}
\tablewidth{0pt}
\tablecaption{Marginalized $1\sigma$ Errors on Dark Energy Parameters 
with All the Other Cosmological Parameters Floating\label{tab:dec}}
\tablehead{\colhead{Probe} & \colhead{$D_i$} & \colhead{$G_i$} &
\colhead{$w_0$} & \colhead{$w_{a}$} & \colhead{$w_{p}$}&
\colhead{EP$^{-1}$}}
\startdata
WL & Y & Y & 0.078        & 0.22    & 0.023    & \phd200 \\ 
\cline{2-7} \\[-1.8ex]
   & Y & M & 0.23\phn     & 0.71    & 0.054    & \phd\phn26 \\
   & M & Y & 1.4\phn\phn  & 2.8\phn & 0.23\phn & \phn1.6 \\ 
\hline \\[-1.5ex]
BAO & Y & Y & 0.33\phn    & 0.81    & 0.044    & \phd\phn28\\ 
\cline{2-7} \\[-1.8ex]
    & Y & M & 0.34\phn    & 0.85    & 0.045    & \phd\phn26 \\
    & M & Y & 3.1\phn\phn & 5.5\phn & 0.87\phn & 0.21
\enddata
\tablecomments{
``Y'' indicates that the set of parameters are 
projected onto cosmological parameter space, and ``M'' indicates that 
they are marginalized. Weak priors are applied to cosmological 
parameters except $w_0$ and 
$w_a$, as discussed in Section \ref{sec:dgcon}. For both BAO and WL, 
the distance is far more powerful than the growth factor in constraining
$w_0$ and $w_a$.}
\end{deluxetable}

\chgd{The parameter splitting technique has been applied to CMB, 
galaxy clustering, SN, and WL data to check the consistency of dark 
energy models \citep{wangs07}. The results do show that the error on
$w_g$ is several times larger than that on $w_d$, even without the 
galaxy clustering and SN data that provide good distance measurements 
at low redshift.}

Another approach to isolate the roles played by the distance and 
growth factor is to separately marginalize 
distance or growth parameters and evaluate the dark energy constraints 
from the rest. Tables \ref{tab:dec} (with all the other cosmological 
parameters floating) and \ref{tab:decfc} (with all the other 
cosmological parameters fixed) present such tests. 
There is no doubt that the growth factor is important for probing dark 
energy with WL tomography, as, in both tables, the EP increases by 
a factor of $\sim 8$ if the growth parameters are marginalized. 
However, given the increase of EP by a factor of $\sim 100$ when the
distance parameters are 
marginalized, the growth factor is clearly much less constraining on 
the dark energy EOS than the distance.

\begin{deluxetable}{l l l c c c c}
\tablewidth{0pt}
\tablecaption{Marginalized $1\sigma$ Errors on Dark Energy Parameters 
with All the Other Cosmological Parameters Fixed \label{tab:decfc}}
\tablehead{\colhead{Probe} & \colhead{$D_i$} & \colhead{$G_i$} &
\colhead{$w_0$} & \colhead{$w_{a}$} & \colhead{$w_{p}$}&
\colhead{EP$^{-1}$}}
\startdata
WL & Y & Y & 0.062    & 0.15    & 0.0069    & \phd990 \\ 
\cline{2-7} \\[-1.8ex]
   & Y & M & 0.12\phn & 0.32    & 0.029\phn & \phd110 \\
   & M & Y & 0.84\phn & 1.8\phn & 0.036\phn & \phd\phn15 \\
\hline \\[-1.5ex]
BAO & Y & Y & 0.19\phn    & 0.39    & 0.022\phn & \phd120 \\ 
\cline{2-7} \\[-1.8ex]
    & Y & M & 0.20\phn    & 0.42    & 0.022\phn & \phd110 \\
    & M & Y & 1.3\phn\phn & 2.3\phn & 0.072\phn & \phn6.2 
\enddata
\tablecomments{Same as Table \ref{tab:dec} but with all the other 
cosmological parameters fixed.}
\end{deluxetable}

\begin{figure*}
\centering
\epsscale{1.13}
\plotone{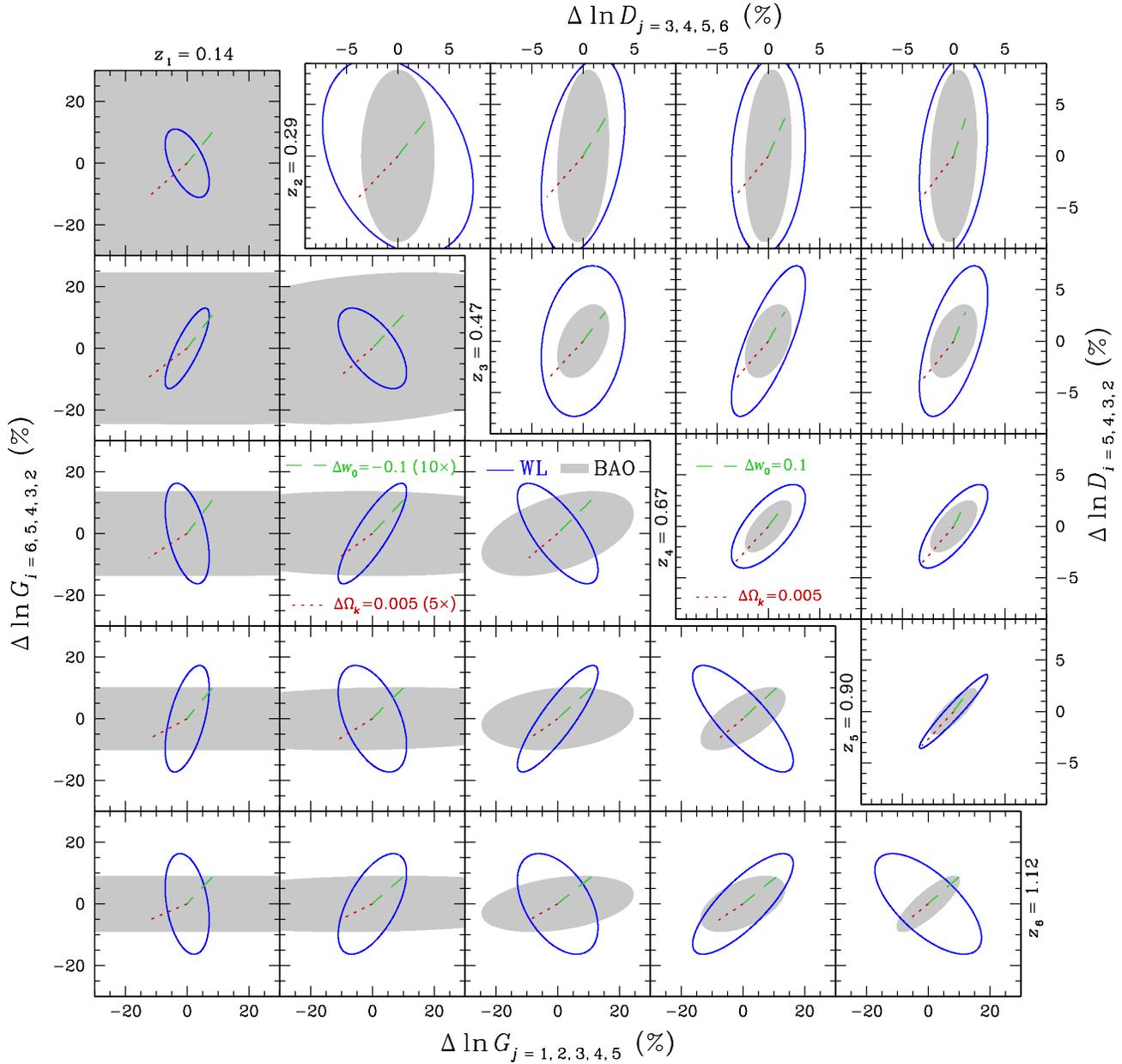}
\caption{Comparison vs. redshift of WL (solid lines) and BAO (shaded 
areas) constraints on the distance (upper triangle) and growth (lower 
triangle) parameters and sensitivities of $\ln D_i$ and $\ln G_i$ 
to $w_0$ and $\Omega_k$.
The error ellipses are $1\sigma$ and marginalized over all other 
parameters including irrelevant $D_i$s and $G_i$s.
The line segments correspond to the changes, 
$\Delta \ln G_i$--$\Delta \ln G_j$ and 
$\Delta \ln D_i$--$\Delta \ln D_j$,
due to small perturbations $\Delta w_0 = \pm 0.1$ (dashed lines) and 
$\Delta \Omega_k = 0.005$ (dotted lines). 
Note that $\Delta \ln G_i$--$\Delta \ln G_j$ are enlarged 10 times for 
$\Delta w_0$ and 5 times for $\Delta \Omega_k$.
Though not shown, the amount of change, $\Delta \ln G_i$ 
($\Delta \ln D_i$), with $\Delta w_a = \pm 0.1$ is 
roughly half (one third) of that with $\Delta w_0 = \pm 0.1$. 
The WL growth constraints are tighter at lower redshift where the 
parameter sensitivities are higher. Moreover, the WL growth error 
ellipses are often oriented in the most sensitive direction to the
parameters, so that the WL growth constraints provide more dark energy
information than the BAO ones (Tables~\ref{tab:dec} \& \ref{tab:decfc}), 
even when the marginalized BAO growth errors are smaller 
(Fig.~\ref{fig:dgcon}).
\label{fig:dge}}
\end{figure*}

Tables \ref{tab:dec} and \ref{tab:decfc} include BAO results as well.
We see in both tables that the dark energy constraints are weakened 
by less than $10\%$ when the growth parameters are marginalized but 
more than an order of magnitude when the distance parameters are 
marginalized. This demonstrates that the
growth factor plays almost no role in constraining $w_0$ and $w_a$ with
\phz{} BAO, even if it can be measured from the shape
of the nonlinear power spectrum (Fig.~\ref{fig:dgcon}).

Comparing the separate BAO and WL constraints from $D_i$ in Tables 
\ref{tab:dec} and \ref{tab:decfc}, we find that for our fiducial survey,
BAO and WL achieve similar EPs (though different
errors on $w_0$ and $w_a$ individually) 
in the absence of the growth information (and systematics). More
interestingly, WL $G_i$ provides more information on the
dark energy EOS than BAO $G_i$, despite that the errors of the latter 
are smaller than those of the former at $z>0.7$. The reason is that
the growth factor is less sensitive to $w_0$ and $w_a$ at higher 
redshift and that the low-redshift growth factors, which WL 
measures better than BAO, are helpful in breaking
the degeneracy between $\Omega_k$ and $w_a$ (as are high-redshift 
distances) (\citealt{zhan06d}; ZK06b).

To gain a further understanding of the roles of the distance and growth 
factor in constraining dark energy, we show in Fig.~\ref{fig:dge} 
marginalized $1\sigma$ contours of $\Delta\ln G_i$--$\Delta\ln G_j$ 
(lower triangle) and $\Delta\ln D_i$--$\Delta\ln D_j$ (upper triangle) 
for WL (solid lines) and 
BAO (shaded areas). The line segments in the figure indicate
the amount of change, $\Delta\ln G_i$ and $\Delta \ln D_i$, due to 
small perturbations $\Delta w_0 = \pm 0.1$ (dashed lines) and 
$\Delta \Omega_k = 0.005$ (dotted lines). The sensitivities of 
$\ln G_i$ and $\ln D_i$ to $w_a$ (not shown) are roughly 
$1/2$ and $1/3$ of those to $w_0$, respectively. From these line 
segments, one can see that the sensitivity of the growth factor to 
$\Omega_k$ and that
of the distance to $w_0$ (and $w_a$ as well) decrease quickly with 
increasing redshift. The sensitivity of the growth factor to $w_0$ and 
$w_a$ has a shallow peak around $z \sim 0.6$.

Fig.~\ref{fig:dge} illustrates a major advantage of WL growth factors 
over BAO growth factors, if ever measured, in constraining the dark 
energy EOS: WL determines the growth factor more accurately than 
BAO at low redshifts where parameter sensitivities are higher. 
Another prominent difference between the BAO and WL constraints on
the growth factor is that the WL $\Delta \ln G_i$--$\Delta \ln G_j$ 
contours are often oriented in the most sensitive direction to the 
parameters, 
i.e., orthogonal to $\Delta \ln G_i$--$\Delta \ln G_j$ caused by 
$\Delta w_0$ or $\Delta \Omega_k$, whereas the BAO ones are generally 
in the least sensitive direction. Therefore, one can still extract 
more information from the WL $G_i$ even when the marginalized errors
of BAO $G_i$ are smaller. Such an advantage is seen up to $z \sim 2$
beyond which the WL contours start to enclose the BAO contours 
completely, but the parameter sensitivities also become fairly low
at $z > 2$.

\subsection{Insights from Distance Eigenmodes} \label{sec:deigen}

Fig.~\ref{fig:dgcon} and Tables \ref{tab:dec} and \ref{tab:decfc} 
pose another puzzle: while the BAO distances are more accurate than
the WL distances, the errors on $w_0$ and $w_a$ from the BAO distances
are actually larger than those from the WL ones. ZK06b postulates that 
the correlation of the WL distance errors might hold the answer, but,  
unlike the growth factor error contours, the WL and BAO distance
error contours 
in Fig.~\ref{fig:dge} are generally in the same direction with 
the former enclosing the latter. To resolve this puzzle, one has to
realize that the contours in Fig.~\ref{fig:dge} are marginalized 
over all the other parameters including irrelevant distance and growth 
parameters, and useful information may have been lost in the process. 
To examine such information, an eigenmode analysis is needed.

Let $\bit{C}_{\rm D}$ be a distance covariance (sub-)matrix from 
Section~\ref{sec:dgcon}. By definition, it is marginalized over all
the other parameters with their priors.
The distance eigenmodes $\bit{E}_I = (E_{I1}, E_{I2},\ldots,
E_{I, N_p-1})^{\rm T}$ satisfy
\begin{equation} \label{eq:evec}
\bit{C}_{\rm D} \bit{E}_I = \lambda_I \bit{E}_I
\quad I = 1, \ldots , N_p-1, 
\end{equation}
where $\lambda_I$ is the eigenvalue corresponding to $\bit{E}_I$,
$N_p - 1$ is the number of distance parameters (remember
that $D_0\equiv 0$ is replaced by $H_0$), and capitalized indices 
are used to avoid confusion with the 
indices of the distance parameters. For convenience, we sort  
$\bit{E}_I$ by their eigenvalues in ascending order.
Since the eigenmodes are orthonormal to each other, their covariance
matrix, $\boldsymbol{\Lambda}$, is diagonal with the variance 
$\sigma^2(\bit{E}_I) = \lambda_I$, i.e.,
\begin{equation} \label{eq:covE}
\Lambda_{IJ} = \delta_{IJ}^{\rm K} \lambda_I.
\end{equation}
One can interpret these $\bit{E}_I$ as the modes of departure of the 
distances from the fiducial model and $\lambda_I$ as a measure of how 
well such deviations can be determined, regardless the driving mechanism.

With equations (\ref{eq:evec}) and (\ref{eq:covE}), we project the 
marginalized distance Fisher matrix $\bit{C}_{\rm D}^{-1}$ into 
$w_0$--$w_a$ space
\begin{eqnarray} \nonumber
F_{\alpha\beta} &=& \sum_{i,j = 1}^{N_p-1}
\frac{\partial \ln D_i}{\partial p_\alpha} 
\left(\bit{C}_{\rm D}^{-1}\right)_{ij}
\frac{\partial \ln D_j}{\partial p_\beta} \\ 
&=& \sum_{i,j,I = 1}^{N_p-1}\frac{\partial\ln D_i}{\partial p_\alpha}
E_{Ii} \lambda_I^{-1} E_{Ij}
\frac{\partial\ln D_j}{\partial p_\beta},
\end{eqnarray}
with $\bit{p}=(w_0,w_a)^{\rm T}$.
We see that the contributions of the distance
eigenmodes to the final Fisher matrix are separable
\begin{equation} \label{eq:FI}
F_{\alpha\beta}= \sum_{I=1}^{N_p-1} \frac{\mathcal{E}_{I\alpha}
\mathcal{E}_{I\beta}}{\lambda_I},
\end{equation}
where $\mathcal{E}_{I\alpha} = \sum_i E_{Ii} \partial \ln D_i /
\partial p_\alpha$.
In other words, the $w_0$--$w_a$ Fisher matrix consists of 
statistically independent contributions from the distance eigenmodes.

The number of distance parameters, $N_p - 1$, does 
not have much impact on the projected dark energy constraints, but it
can assert artificial restrictions on these modes. Thus,
before drawing conclusions from the distance eigenmodes, one must
ensure that the interpolation has sufficient degrees of freedom
to reflect the intrinsic properties of the distance eigenmodes of 
the dark energy probes, though having too many distance parameters 
will leave the results prone to numerical errors. 
Since the intrinsic eigenmodes do not vary with $N_p$, 
one can increase $N_p$ until the shapes of the 
eigenmodes converge. In addition, the variance $\sigma^2(\bit{E}_I)$ 
should grow in proportion with the degrees of freedom, $N_p - 1$, once
the convergence is reached. By the same token, we change the priors
on $\ln D_i$, $\ln G_i$, and $\ln b_i$ to $100 N_p^{1/2}$, though 
these priors are already too weak to make any difference.

\begin{figure}
\centering
\epsscale{1}
\plotone{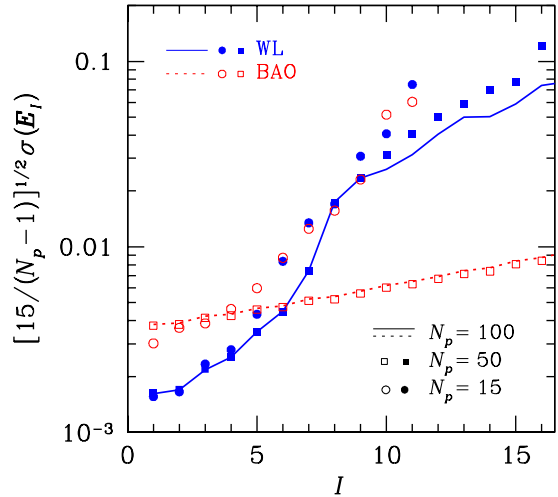}
\caption{Comparison of the errors of the distance eigenmodes, 
$\sigma(\bit{E}_I)$, from WL (solid line and filled symbols) and 
from BAO (dotted line and open symbols). All the other parameters 
are marginalized, and the errors are scaled by the square root of 
the number of distance parameters, $N_p - 1$, with $N_p = 15$ 
(circles), 50 (squares), and 100 (lines). The WL distances have 
several eigenmodes that are better determined than all the BAO 
distance eigenmodes. The alignment between the $N_p = 50$ and 
$N_p = 100$ results suggests that the best determined modes have
converged at $N_p \gtrsim 50$.
\label{fig:eval}}
\end{figure}

Fig.~\ref{fig:eval} shows the errors of the distance eigenmodes scaled 
by $(N_p-1)^{-1/2}$ for WL (solid line and filled symbols) and BAO 
(dotted line and open symbols) with $N_p = 15$ (circles), 
50 (squares), and 100 (lines). The alignment between the $N_p=50$ 
and $N_p = 100$ eigenmode errors suggests that $N_p \gtrsim 50$ is 
sufficient for the best determined modes to converge, which is 
confirmed upon inspection of the modes. Thereafter, we set
$N_p = 100$.

An interesting result in Fig.~\ref{fig:eval} is that some WL distance 
eigenmodes are better determined than all the BAO distance 
eigenmodes, even though the marginalized errors of individual WL 
distance parameters are all larger than those of the BAO distance
parameters in Fig.~\ref{fig:dgcon}. This is predicted in ZK06b to be
an advantage of the WL distance measurements, albeit that the
best determined modes do not necessarily provide the most information
on dark energy.

\begin{figure*}
\centering
\epsscale{1}
\plottwo{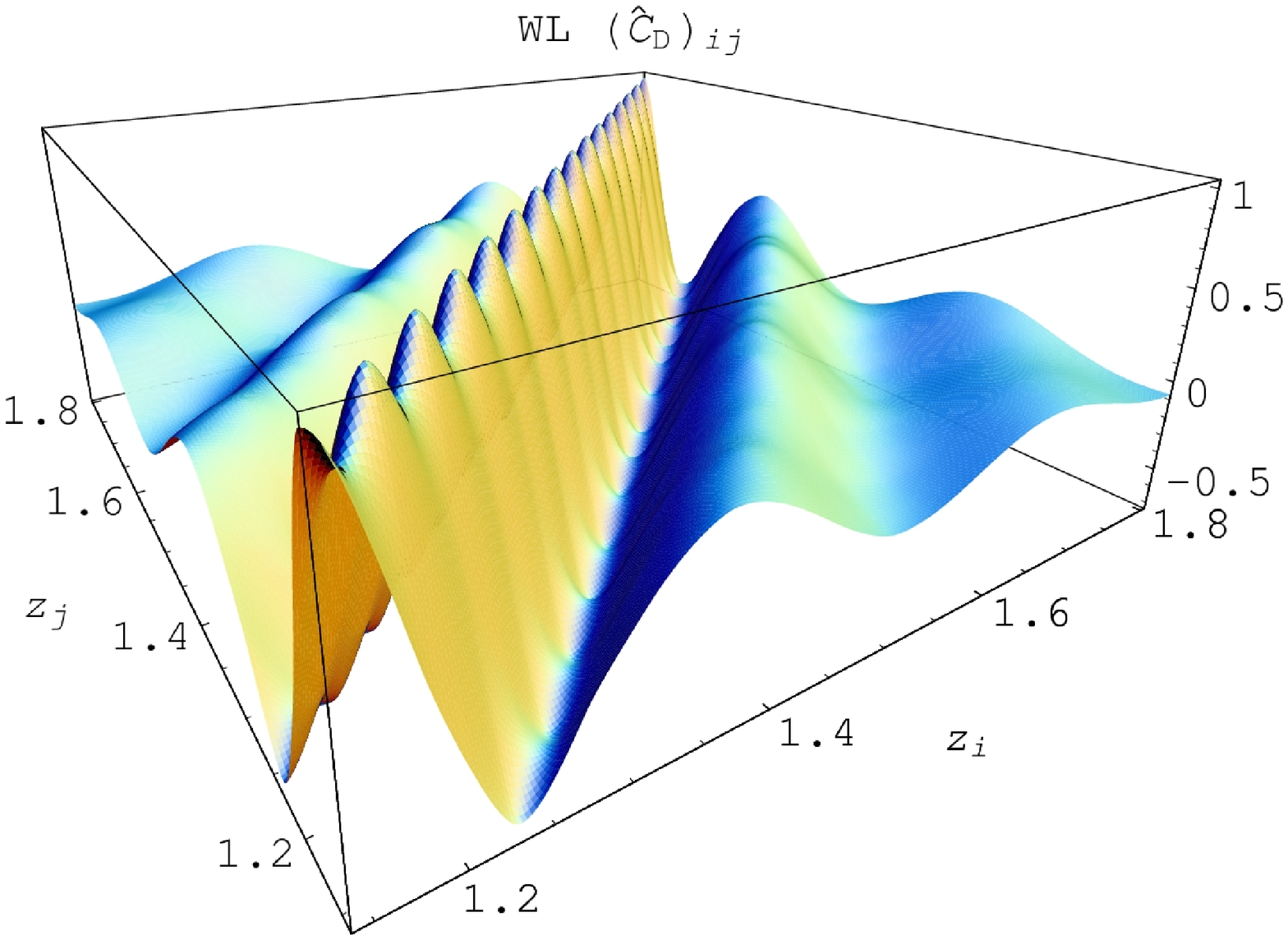}{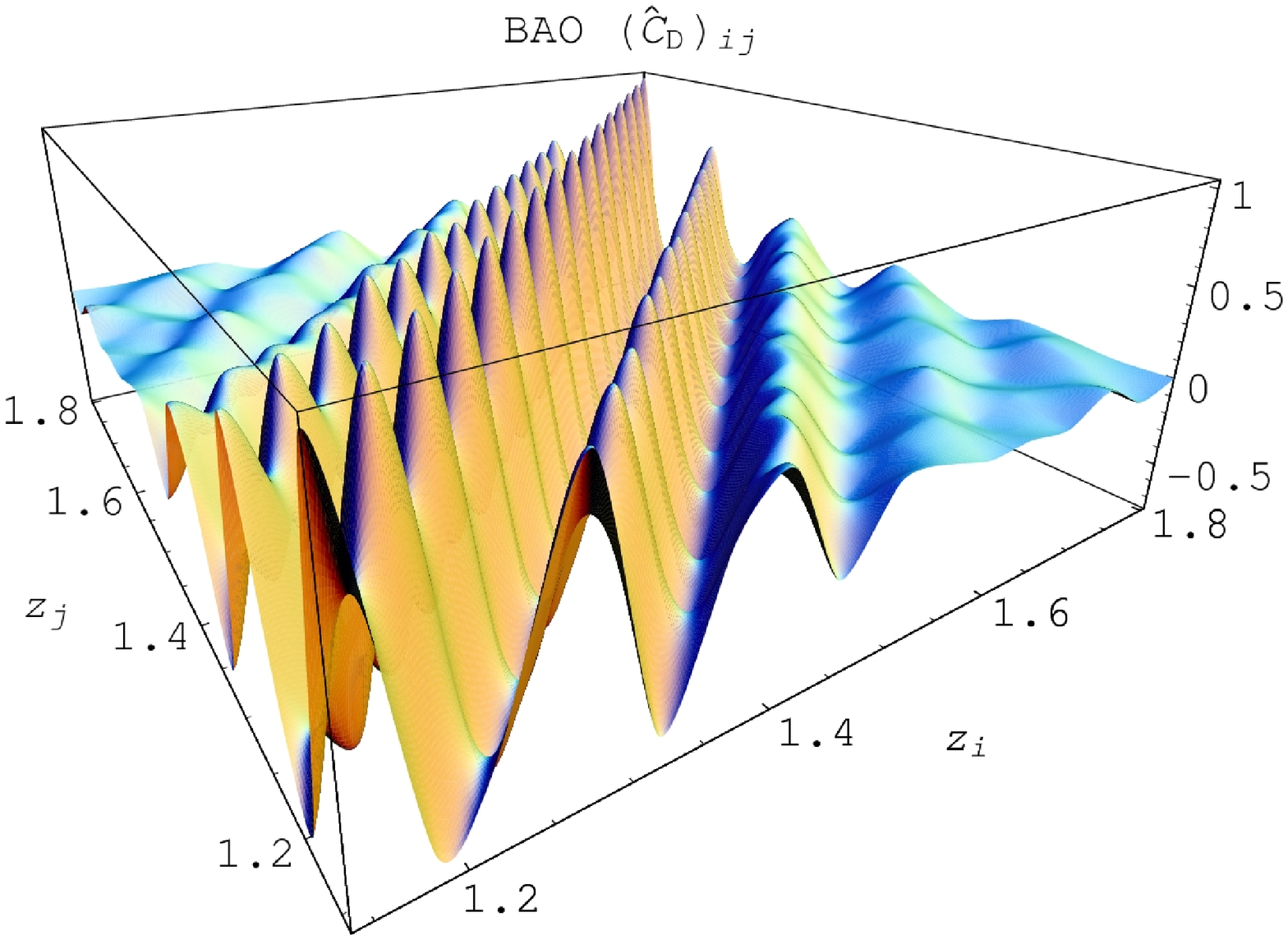}
\caption[f5]{Structure of the distance covariance matrix from WL
(left panel) and that from BAO (right panel). The difference between
the two covariances reflects the complementarity of BAO and WL 
distances. Elements of the 
covariance matrices are scaled by the geometric mean of corresponding 
diagonal elements. Only 16 distance parameters between $z = 1.14$ and 
$1.81$ are presented here. The covariance matrices have been smoothly 
interpolated to show the underlying structure, and the discrete matrix
elements are identifiable with the peaks and troughs if present. 
\chgd{Note that the values of the matrix elements change rather
slowly along the diagonal direction, so the apparent oscillations 
along that direction are not real. They are the result of isotropic 
rendering by the graphics software for a grid of data that vary more
rapidly in the direction perpendicular to the diagonal than in the 
direction of the diagonal.}
\label{fig:dcov}}
\end{figure*}

Fig.~\ref{fig:dcov} shows a small piece of the distance
covariance matrix of WL (left panel) and that of BAO (right panel). 
The covariance matrices are normalized
\[ 
(\hat{C}_{\rm D})_{ij} = \frac{(C_{\rm D})_{ij}}
{\sqrt{(C_{\rm D})_{ii}(C_{\rm D})_{jj}}},
\]
so that $|(\hat{C}_{\rm D})_{ij}| \le 1$. Because the lensing kernel 
is much broader than the galaxy power spectrum kernel, we expect the 
WL distance covariance to vary more slowly than the BAO distance 
covariance, which is indeed seen in Fig.~\ref{fig:dcov}. The features
of the distance covariance matrices in Fig.~\ref{fig:dcov} are typical 
over the whole redshift range, with some broadening at both low-$z$ 
and high-$z$ ends, though the values of the individual covariance 
elements do vary with the details of the calculations such as the 
interpolation scheme and tomographic binning.

\begin{figure*}
\centering
\epsscale{1}
\plotone{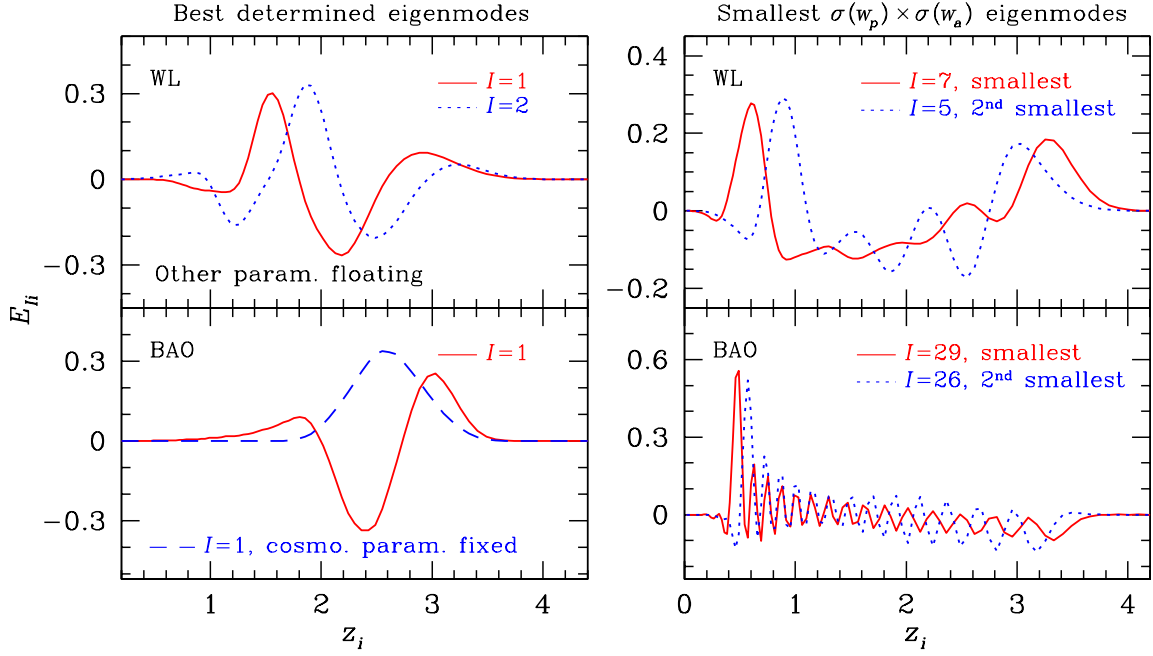}
\caption{The best determined distance eigenmodes (left panels)
and the ones that best constrain $w_0$ and $w_a$ (right panels) 
from WL (upper panels) and BAO (lower panels). The modes are sorted
by their error in an ascending order.
All the other parameters are floating, except that one of the best 
determined BAO distance eigenmodes (dashed line) is obtained with all 
the cosmological parameters fixed. The best determined WL 
distance eigenmodes do not change much whether the cosmological 
parameters are fixed or not. While WL is best at measuring distance
modes that are insensitive to the absolute distance normalization
(upper left panel), its constraints on $w_0$ and $w_a$ are derived 
largely from modes that are sensitive to the distance normalization
(upper right panel). 
\label{fig:evec}}
\end{figure*}

The characteristic difference between the distance
covariance matrices of the two techniques must hold the answer to the 
paradox brought up at the beginning of this subsection: although BAO
determines the distance more accurately than WL (the WL distance 
error contours also enclose the BAO ones), the EPs of each
technique, when the growth parameters are marginalized, are the same 
with WL having smaller marginalized errors on $w_0$ and $w_a$. 
Moreover, this difference also suggests that BAO and WL distance 
measurements are complementary.

We present several eigenmodes of BAO and WL distances in  
Fig.~\ref{fig:evec}. While not shown, the best determined WL modes 
(upper left panel)
do not change much whether the cosmological parameters are fixed or 
not. These modes have close to zero mean, i.e., $|\sum_i E_{Ii}| \ll 1$,
and, hence, are not sensitive to the absolute normalization of the 
distance. To see this, assume that all the distances are subject to 
the same multiplicative factor $1+f$ ($|f|\ll 1$), so that the 
fractional errors $\Delta \ln D_i = f$. After projecting the distance
errors on the mode $\bit{E}_I$, one gets the amplitude 
$\sum_i \Delta \ln D_i E_{Ii} = f \sum_i E_{Ii}$, i.e., 
$|\sum_i E_{Ii}|$ gauges the sensitivity of an eigenmode to a change 
in the distance normalization. Thus the best determined WL modes show 
that WL is best at measuring the relative distance, which agrees
with the expectation that WL tomography measures distance ratios from
the lensing kernel. 

\begin{figure*}
\centering
\epsscale{0.86}
\plotone{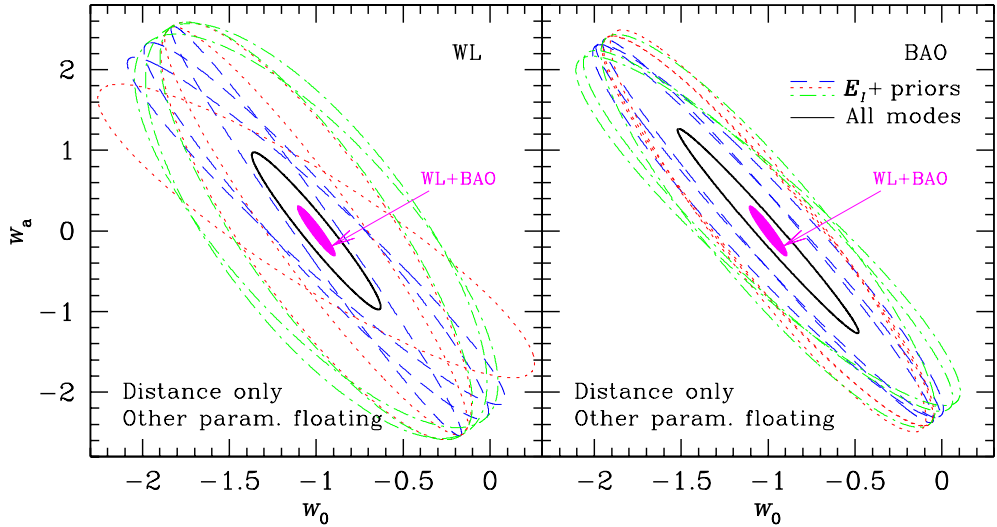}
\caption{Decomposition of the $w_0$ and $w_a$ constraints from the
smallest-EP distance eigenmodes (broken lines) for WL (left panel) and 
BAO (right panel). The WL distance eigenmodes are fairly complementary
to each other, so that only 5 modes are needed to account for nearly 
all the dark energy constraints of the 99 distance eigenmodes (solid
line). The BAO modes generally produce narrower $w_0$--$w_a$ error 
contours than WL ones, but their orientations are similar. This 
makes the BAO modes less effective in reducing the marginalized errors
on $w_0$ and $w_a$ individually, even though the error contours from
all the distance modes (solid lines) of LSST BAO and WL have similar 
areas (or EPs, see Table~\ref{tab:dec}). BAO and WL are highly 
complementary to each other, and the joint analysis (shaded area in 
both panels) reduces the EP by a factor of $\sim 9$ with respect to 
the EP of either technique alone.
We have applied the priors $\sigma_P(w_0)=1$ and $\sigma_P(w_{a})=2$ 
to individual modes and marginalized over all the other parameters 
with weak priors. The indices of the WL smallest-EP modes are 
$I = 7, 5, 17, 12, 8, 4, 14, 15, 23$, in an order of increasing 
error ellipse area (or EP), and those of BAO modes are 
$I = 29, 26, 27, 39, 31, 35, 38, 32, 23$. The fact that the best 
determined distance eigenmodes from both BAO and WL are not the most 
sensitive modes to $w_0$ and $w_a$ means that the two dark energy 
probes can potentially explore a larger dark energy parameter space 
and afford some redundancy to control certain systematic uncertainties.
\label{fig:evw}}
\end{figure*}

The best determined BAO distance eigenmodes (lower left panel in
Fig.~\ref{fig:evec}) display more sensitivity to the absolute distance
than the best determined WL distance eigenmodes, 
especially when the standard ruler of BAOs is precisely known by 
fixing the cosmological parameters. The best determined BAO modes 
peak at higher redshift than the WL ones
because of the following reasons. First, the severe truncations of 
multipoles of the galaxy power spectra at low redshift, e.g., 
$\ell_{\rm max}=732$ at $z_p=1.05$ vs. $\ell_{\rm max}=3000$ at 
$z_p=2.09$, increase the low-$z$ BAO distance 
errors. Second, given the same source galaxy distribution, the lensing 
kernel peaks half-way between the observer and the source, whereas the 
galaxy power spectrum kernel peaks at the source. Finally, at higher 
redshift, the BAO features move to higher multipoles, where the smaller 
cosmic variance leads to better measurement of the distance.

The best determined distance eigenmodes are not necessarily the ones
that contribute the most to the constraints on the dark energy EOS,
because they may not be sensitive to a particular deviation of the 
distance--redshift relation caused by dark energy.
Phrased another way, \citet{deputter07} point out that the eigenmode 
analysis gives the expected noise of the measurements, i.e., the 
error of the eigenmode, but the signal, i.e., the amplitude of the 
mode, and the signal-to-noise ratio are not known unless comparison 
models are assumed \citep[e.g.,][]{barnard08} or until  
measurements are made. 

For the $w_0$--$w_a$ parametrization of the 
dark energy EOS, the comparison models populate 
the $w_0$--$w_a$ plane with the fiducial model (cosmological constant)
at $w_0 = -1$ and $w_a = 0$. The contribution of each distance 
eigenmode to the $w_0$--$w_a$ Fisher matrix is statistically 
independent of each other, so that we can evaluate the usefulness of 
an eigenmode by its constraint on $w_0$ and $w_a$. Since a single 
mode cannot constrain both $w_0$ and $w_a$, we apply the priors 
$\sigma_P(w_0) = 1$ and $\sigma_P(w_a) = 2$ in the projection of 
$\sigma(\bit{E}_I)$ onto $w_0$ and $w_a$ via equation (\ref{eq:FI}). 

For a broad class of dark energy models, their effect on the distance
is concentrated at low redshift without rapid oscillations. This 
means that a change of sign in $E_{Ii}$ at low redshift will result 
in a reduction of the sensitivity to the dark energy EOS. Therefore,
dark-energy-efficient modes should be roughly those with a large 
value of $|\sum_i E_{Ii}|$ at low redshift. The right column 
of Fig.~\ref{fig:evec} shows the distance eigenmodes with the smallest 
EPs for WL (upper panel) and BAO (lower panel). As expected, they all 
have a major peak at $z \lesssim 1$. The  BAO modes (lower right
panel) exhibit oscillatory behavior at $z \gtrsim 0.6$ because they are 
orthogonalized to other better determined modes that vary smoothly.

The fact that the best determined distance eigenmodes (with other 
parameters floating) from both WL 
and BAO are not the most sensitive modes to $w_0$ and $w_a$ means that
the two dark energy probes can potentially explore a larger dark energy
parameter space \citep[e.g.,][]{albrecht07, barnard08} and afford some 
redundancy to control certain systematic uncertainties. We also note 
that although $\sigma(\bit{E}_I)$ and EP do not have one-to-one
correspondence, contributions to the dark energy constraints naturally
come mostly from the best determined modes. For WL, the 
dark-energy-sensitive modes are 
confined within $I \lesssim 20$, whereas for BAO, they are spread out 
to $I \lesssim 40$. This is reflected in the indices of the 
smallest-EP modes in Fig.~\ref{fig:evec} as well.

There has been a notion that WL tomography constrains dark energy 
with relative distances or distance ratios from the lensing kernel. 
It is true that the best determined WL distance eigenmodes are 
not sensitive to the normalization of the distance, but the modes that 
contribute the most to the constraints on $w_0$ and $w_a$ are more 
sensitive to the normalization of the distance. 
For example, the average value
of $|\sum_i E_{Ii}|$ of the 5 best determined WL distance eigenmodes
is 0.0094, whereas that of the 5 
smallest-EP modes is 1.7 (for BAO, the values are 0.033 and 1.4, 
respectively). Since the above results are marginalized 
over the growth parameters, which weakens WL's sensitivity to the
absolute distance, it is reasonable to attribute the WL constrains on 
dark energy to its measurements of absolute distances.

To see why the WL distances give smaller errors on $w_0$ and $w_a$,
even though they are less accurately determined than the BAO distances, 
we show in Fig.~\ref{fig:evw} marginalized $1\sigma$ $w_0$--$w_a$ 
error contours of the eigenmodes that have the smallest EPs of each
technique. The priors $\sigma_P(w_0) = 1$ and $\sigma_P(w_a) = 2$ are 
applied, and all the other parameters are marginalized with weak priors. 
The solid error contour in each panel shows the constraints of all
99 distance eigenmodes for each technique. Since the dark energy 
constraint from each eigenmode is independent of those from
other eigenmodes, one can see immediately that 5 WL distance eigenmodes
can account for nearly all the $w_0$ and $w_a$ constraints of WL 
distances. This echoes the conclusion from Fig.~\ref{fig:evec} that 
WL tomography has the potential of constraining more than $w_0$ and 
$w_a$ with all the well determined modes.
Furthermore, the error ellipses of these 5 eigenmodes are oriented 
differently, so that they are effective in reducing the marginalized 
errors on $w_0$ and $w_a$.
The BAO error ellipses are narrower than the WL ones in general, but 
they are aligned in nearly the same direction. As such, combinations
of the \phz{} BAO distance eigenmodes are not effective in reducing 
the marginalized errors of $w_0$ and $w_a$, and many more modes are
needed to match the combined EP with that of WL (Tables~\ref{tab:dec}
and \ref{tab:decfc}). 
Because BAO and WL have different parameter degeneracy directions and 
because the galaxy--shear cross power spectra provide additional 
information, the two techniques are highly complementary to each 
other. Indeed, the joint BAO and WL distance constraints on $w_0$ 
and $w_a$ (shaded area) is roughly 9 times tighter than those from 
either technique alone \citep[see also][]{zhan06d}.

\section{Conclusions} \label{sec:con}

We have disentangled the roles of the distance and growth factor in 
constraining the dark energy EOS and given a detailed comparison 
between the distance measurements from BAO and WL as well as their 
constraints on $w_0$ and $w_a$. We find that if the growth parameters 
are marginalized without being projected onto $w_0$ and $w_a$ 
constraints, the EP of LSST WL will increase by a factor of $\sim 8$.
However, if the distance parameters are marginalized instead, the 
degradation will be two orders of magnitude. This clearly shows that 
the growth factor is important to WL, but the distance is far more 
important. In contrast, the BAO EP increases by only 10\%, when the 
growth parameters are marginalized. It is also interesting that LSST  
BAO and WL achieve similar EPs when the growth parameters are 
marginalized.

The reconstruction of a continuous function depends on how the 
function is represented by discrete parameters. To explore the 
intrinsic properties of the reconstructed distances 
from BAO and WL, we assign a large number of distance and growth 
parameters and examine various aspects of the distance eigenmodes.
This exercise leads to the finding that the WL distance eigenmodes
are more complementary to each other than the BAO ones. As such,
5 WL eigenmodes can provide most of the LSST WL distance constraints 
on $w_0$ and $w_a$, whereas many more are needed for BAO. 
A more useful result is that both LSST BAO and WL have some well 
determined distance eigenmodes that are not very sensitive 
to $w_0$ and $w_a$. These modes may be able to constrain additional 
dark energy parameters and some systematic uncertainties. 

\begin{figure}
\centering
\epsscale{1}
\plotone{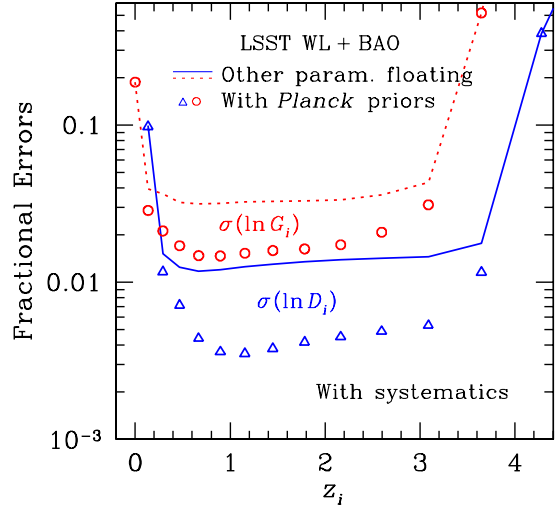}
\caption{Same as Fig.~\ref{fig:dgcon}, but for distance (solid line 
and open triangles) and growth factor (dotted line and open circles) 
constraints from the joint analysis of LSST BAO and WL with a 
conservative level of systematic uncertainties in the photo-z error 
distribution and additive and multiplicative errors in the shear and 
galaxy power spectra (see text for more details). The lines 
represent the results with other parameters floating, and the symbols 
are with CMB priors from \emph{Planck}. The joint constraints on 
distance are relatively insensitive to the assumed systematics. 
\label{fig:dges}}
\end{figure}

The insight gained from the results in Sections~\ref{sec:dgch} and 
\ref{sec:pde} in the absence of various systematics is valuable for 
understanding the techniques. It is also of great interest for future
surveys to have more realistic error estimates that include those 
systematics. To this end, we present in Fig.~\ref{fig:dges} 
constraints on the distance (solid line and open triangles) and 
growth factor (dotted line and open circles) from the joint analysis 
of LSST BAO and WL 
with a conservative level of systematic uncertainties in the photo-z 
error distribution and additive and multiplicative errors in the shear 
and galaxy power spectra. The lines represent the results with other 
parameters floating, and the symbols are with CMB priors from 
\emph{Planck}. By comparing with Fig.~\ref{fig:dgcon}, one sees that 
the joint constraints on distance are relatively insensitive to the 
assumed systematics whereas those on the growth factor are affected 
more.

The results in Fig.~\ref{fig:dges} are marginalized over additional 
30 \phz{} bias parameters $\delta z_i$, 30 \phz{} rms parameters 
$\sigma_{z,i}$, 10 shear additive error parameters $A_j^\gamma$, 
10 shear multiplicative error parameters $f_j^\gamma$, and
20 galaxy additive noise parameters $A_k^\mathrm{g}$. \chgd{
We extend the additive and multiplicative shear power spectrum errors 
in \citet{huterer06} to include the galaxy power spectrum errors:
\begin{eqnarray} \nonumber
(\bit{C}_\ell^{XY})_{ij} &=& (1 + \delta_{X\gamma}^{\rm K} f_i^X + 
\delta_{Y\gamma}^{\rm K} f_j^Y) P_{ij}^{XY}(\ell) + \\
&& \delta_{XY}^{\rm K} \left[\delta_{ij}^{\rm K} X_{\rm rms}^2 
\bar{n}_i^{-1} + \rho^X A_i^X A_j^Y
\left(\frac{\ell}{\ell_*^X}\right)^{\eta^X}\right], \nonumber
\end{eqnarray}
where $\rho^X$ determines how strongly the additive errors of two 
different bins are correlated, and $\eta^X$ and $\ell_*^X$ account
for the scale dependence of the additive errors.
Note that the multiplicative error of galaxy number density is 
degenerate with the galaxy clustering bias and is hence absorbed
by $b_i$. Below the levels of systematics future surveys aim to 
achieve, the most important aspect of the (shear) additive error is
its amplitude \citep{huterer06}, so we simply fix $\rho^X = 1$
and $\eta^X = 0$. For more comprehensive accounts of the above 
systematic uncertainties, see 
\citet{huterer06, jain06, ma06, zhan06d}.}

We have applied priors
$\sigma_P(\delta z) = 2^{-1/2} \sigma_P(\sigma_z) = 0.01(1+z)$ per
\phz{} parameter interval ($\Delta z = 0.17$) in Fig.~\ref{fig:dges}.
\chgd{For \emph{Gaussian} \phz{} errors with $\sigma_z = 0.05(1+z)$, 
these priors correspond to a calibration requirement of 25 spectra 
per redshift interval of 0.17.}
The constraints do not change appreciably even
if one tightens the priors to $0.003(1+z)$, because the \phz{}
sample--\phz{} sample cross correlations can calibrate the uncertainties
in $\delta z$ and $\sigma_z$ to $10^{-3}$ level \citep{zhan06d}.
Similar results are obtained per redshift bin with spectroscopic
sample--\phz{} sample
cross correlations \citep{newman08}. For the shear multiplicative
error, \citet{massey07a} shows that current methods consistently achieve
better than $2\%$ precision. Allowing for another 5 to 10 years of
development, we are hopeful that the multiplicative errors can be
controlled to half a percent, i.e., $\sigma_P(f_i^\gamma) = 0.005$.
\chgd{For the shear additive error, \citet{paulin-henriksson08} finds 
for a 15,000 deg$^2$ ground survey that it will be well under control 
over relevant scales if the point spread function (PSF) is calibrated 
over 100 stars. LSST will be able to use on average 2 -- 3 stars per 
square arcminute to calibrate the PSF in each exposure, so the shear 
additive error will not have a significant impact on scales larger 
than $\sim 5$ arcminute (or $\ell \lesssim 2000$) even for a single 
exposure. This is supported by a study with Subaru data 
\citep{wittman05}, which shows that the correlation of the residual 
stellar shear after the PSF correction using only 0.9 star per square 
arcminute (out of $\sim 8$ arcmin$^{-2}$ available) is well below the 
cosmic shear signal at arcminute scale for a single 10-second 
exposure and that it is roughly inversely proportional to the number 
of exposures. } In addition to deep exposures in
\emph{u}, \emph{g}, and \emph{y}, LSST plans to take 400 exposures
per sky field in each of its \emph{riz} filters (these will be used for 
WL as well as \phz{}). 
Thus it is reasonable to project that the additive shear error will be
sub-dominant to the statistical errors on relevant scales.
Full simulations of LSST performance are in progress.
To be conservative, we assume that $(A_i^\gamma)^2 = 10^{-9}$,
which is roughly the amplitude of a $z \gtrsim 2$ shear power
spectrum at $\ell \sim 1000$. Finally, we infer from the Sloan Digital
Sky Survey galaxy angular power spectrum \citep{tegmark02} that the
additive galaxy power spectrum error due to extinction, photometry
calibration, and seeing will be at $(A_i^\mathrm{g})^2 = 10^{-8}$ level.

With the assumed systematic uncertainties, we find that the joint
analysis of LSST BAO and WL can achieve $\sim 1.5\%$ precision on
11 distances from $z_2 = 0.29$ to $z_{12} = 3.6$ with weak priors
on cosmological parameters. Strong priors from \emph{Planck} will
reduce the errors to $\sim 0.5\%$ level. The corresponding errors on
the growth factors are $\sim 3.5\%$ from $z_1 = 0.14$ to
$z_{11} = 3.1$ without \emph{Planck} and $\sim 2\%$ with \emph{Planck}.
These estimates are obtained without assuming a specific model for
dark energy or modified gravity, so they are fairly model
independent.

Finally, we mention that the distance and growth factor reconstruction
automatically provides a pure metric measurement of the mean
curvature of the universe. The joint LSST BAO and WL analysis can
achieve $\sigma(\Omega_k) = 0.017$ with the above mentioned
systematic uncertainties and weak priors on the other cosmological
parameters. Though less impressive than other model-dependent forecasts
of $\sigma(\Omega_k) \lesssim 10^{-3}$ \citep{knox06c,zhan06d}, this
estimate does not assume any specific dependence of the distance on
cosmological parameters.

\acknowledgments

HZ thanks Gary Bernstein for helpful comments and questions.
This work was supported by a UC Davis
Academic Federation Innovative Developmental Award, DOE grant
DE-FG02-07ER41505, and a TABASGO Foundation grant. LK was 
supported by NSF award number 0709498.

\appendix

\section{Influence of Cosmological Parameters on the Distance
and Growth Factor Constraints} 
\label{sec:dcldq}

We check the results with only one cosmological parameter fixed at a 
time in order to identify which ones contribute the most to
the difference between the constraints with all the other parameters 
floating and those with all the cosmological parameters fixed 
in Fig.~\ref{fig:dgcon}. For BAO distances, the matter 
density $\omega_{m}$, the baryon density $\omega_{b}$, and 
the curvature parameter $\Omega_k^\prime$ are the most crucial
parameters, and they have comparable impact on the distance errors
with $\Omega_k^\prime$ weighing more at higher redshift. This 
is expected qualitatively because $\omega_{m}$ and 
$\omega_{b}$ determines 
the standard ruler of BAOs and because the fractional difference 
between the comoving angular diameter distance $D_{A}(z)$ and
the comoving distance $D(z)$,
\[
\frac{D_{A}(z)}{D(z)} - 1 \simeq \frac{1}{6}
\frac{\Omega_k^\prime H_0^2}{c^2} D^2(z)\quad 
\mbox{for}\ |\Omega_k^\prime| \ll 1,
\]
is larger at higher redshift. The single most
important parameter for the BAO growth factors (with galaxy bias 
parameters floating) is the normalization of the matter power 
spectrum, $\Delta_{R}^2$. The reason is that fixing 
$\Delta_{R}^2$ removes its degeneracy with each $G_i$ for any
single galaxy power spectrum.

The WL results are less obvious. The matter density 
$\omega_{m}$ affects the WL distance constraints the 
most, then the primordial spectral index $n_{s}$, and then 
the normalization $\Delta_{R}^2$. Fixing these parameters reduces
the uncertainty in the amplitude and shape of the matter power 
spectrum, which is helpful. However, 
the same argument applies to other parameters 
such as the baryon density $\omega_{b}$ and the running of the 
spectral index $\alpha_{s}$ as well, even though $\omega_{b}$
and $\alpha_{s}$ have much less impact on the distance constraints.
Yet more puzzling is that the WL growth constraints hardly change 
whether the cosmological parameters (especially 
$\Delta_{R}^2$) are fixed or not. 

\begin{figure*}
\centering
\epsscale{0.98}
\plotone{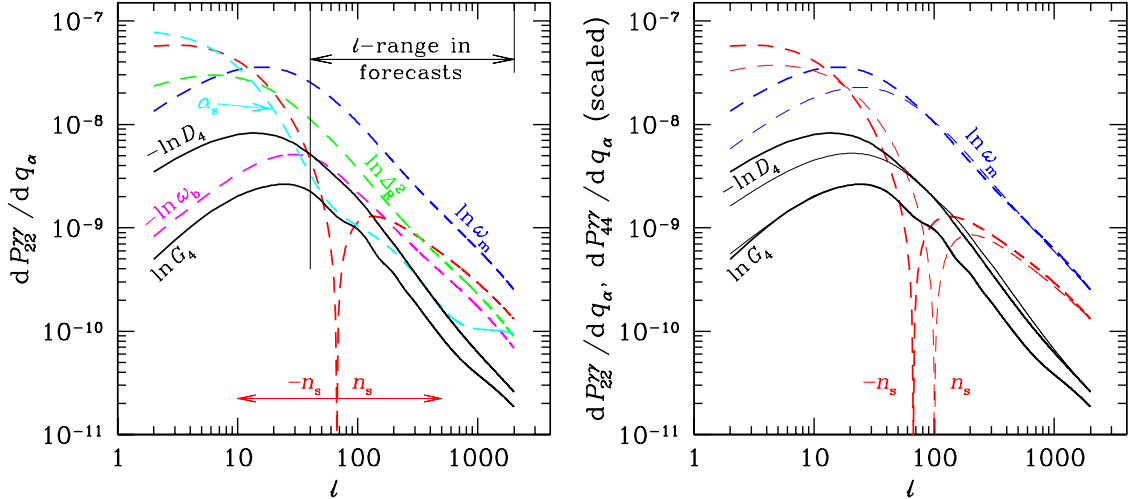}
\caption{\emph{Left panel}: Derivatives of the auto shear power spectrum 
$P_{22}^{\gamma\gamma}(\ell)$ (in the redshift bin $0.7 < z_{p} \le 1.05$)
with respect to a subset of parameters as labeled:
$\ln\omega_{m}$, $-\ln\omega_{b}$, $\pm n_{s}$,
$\alpha_{s}$, and $\Delta_{R}^2$ in dashed lines, 
and $-\ln D_4$ and $\ln G_4$ at $z_4 = 0.67$ in solid lines.
\emph{Right panel}: Comparison between $P_{22}^{\gamma\gamma}(\ell)$ (thick lines) 
and $P_{44}^{\gamma\gamma}(\ell)$ ($1.4 < z_{p} \le 1.75$, thin lines) 
derivatives with respect to $\ln\omega_{m}$, $\pm n_{s}$, 
$-\ln D_4$, and $\ln G_4$. 
To show the difference in the slopes, we normalize 
the $P_{44}^{\gamma\gamma}(\ell)$ derivatives to the corresponding
$P_{22}^{\gamma\gamma}(\ell)$ derivatives at $\ell = 2000$ where the 
statistical errors are roughly at the minimum. The derivatives with
respect to $n_{s}$ reverse sign near $\ell \sim 100$ because 
the matter power spectrum is pivoted at 
$k = 0.05\, \mbox{Mpc}^{-1} = \ell/D_{A}(z)$
for $n_{s}$. The pivot in multipole space
moves toward higher $\ell$ for shear power spectra in (or between)
higher redshift bins. 
\label{fig:dcl}}
\end{figure*}

We do not find an intuitive explanation for the behavior of the WL
distance constraints. If the derivatives of the shear power spectra
with respect to two parameters are nearly degenerate, i.e., roughly 
proportional to each other, then the prior on one parameter will 
affect the constraint on the other parameter significantly. However,
this is not the case; 
as demonstrated in the left panel of Fig.~\ref{fig:dcl}, none of the 
derivatives of the auto shear power spectrum $P_{22}^{\gamma\gamma}(\ell)$
(in the redshift bin $0.7 < z_{p} \le 1.05$) with respect to
$\ln \omega_{m}$, $n_{s}$, and $\ln \Delta_{R}^2$ 
resemble the derivative with respect to $\ln D_4$ ($z_4 = 0.67$).
Adding to the complexity is that the derivatives vary in both shape and 
amplitude from one power spectrum to another (right panel
of Fig.~\ref{fig:dcl}) and that the derivative with respect to one 
distance parameter differs from that with respect to another distance
parameter (not shown). This is true for the normalized derivatives 
$d\ln P_{ij}^{\gamma\gamma}/dq_\alpha$ as well.

The fact that the prior on the power spectrum normalization 
$\Delta_{R}^2$ affects the WL distance errors corroborates the 
conclusion in ZK06b that the WL technique measures the distance 
from the amplitudes of the shear power spectra. 
Closer inspection of equation (\ref{eq:aps}) shows that 
$\omega_{m}$ also alters the amplitude of the shear power 
spectrum. The effect is twofold:
(1) the combined factor $\omega_{m}^2 / a^2$ in the window functions 
scales (up to a constant factor) the overdensity power spectrum 
$\Delta_\delta^2(k)$ to the potential power spectrum 
$\Delta_\phi^2(k)$ that is directly related to lensing, 
and (2) an increasing 
$\omega_{m}$ moves the broadband turn-over of the transfer
function toward small scales, giving a small boost to the shear power 
spectra over the multipole range in our forecasts. This is consistent
with $dP_{22}^{\gamma\gamma}/d\ln\omega_{m} \gtrsim 2 
dP_{22}^{\gamma\gamma}/d\ln\Delta_{R}^2$ at 
$\ell \ge 40$ and, hence, the prior on $\omega_{m}$ having a 
greater impact on the WL distance constraints in Fig.~\ref{fig:dgcon}. 

In the right panel of Fig.~\ref{fig:dcl}, the derivatives of different
shear power spectra with respect to the growth factor parameter $G_4$ 
have almost the same shape. It is likely that this characteristic  
sets apart the growth parameters from the rest, so that the WL growth 
factor constraints are practically not affected by the priors on the 
cosmological parameters in Fig.~\ref{fig:dgcon}. To see why these 
derivatives are nearly proportional to each other, we make the 
approximations $\Delta_{\delta}(k;z) \sim \tilde{G}(z) 
\Delta_{\delta}(k)$
and 
\[
\frac{d \Delta_\delta^2(k;z)}{d G_m}\sim 
\frac{d \tilde{G}^2(z)}{d G_m}\Delta_{\delta}^2(k) \sim
\delta^{\rm D}(z - z_m) 
\Delta_{\delta}^2\left[\frac{\ell}{D_{A}(z)}\right],
\]
where $\tilde{G}(z)$ is the interpolated growth factor, and
$\delta^{\rm D}(z - z_m)$ is the Dirac delta function. We have then 
\begin{equation} \label{eq:dpdg} 
\frac{dP_{ij}^{\gamma\gamma}(\ell)}{d\ln G_m} \propto A_{ij}(z_m)\, 
\ell^{-3} G_m \Delta_{\delta}^2 \left[ \frac{\ell}{D_{A}(z_m)} \right]
\end{equation}
with $A_{ij}(z) = H(z) D_{A}(z) W_i^\gamma(z) W_j^\gamma(z)$
showing that the redshift bins have no influence on the shape of
the derivatives $dP_{ij}^{\gamma\gamma}/dG_m$. 
Since equation (\ref{eq:dpdg})
does not integrate the matter power spectrum over the scales, 
the BAO features become visible in $dP_{ij}^{\gamma\gamma}/dG_4$
at $\ell \sim 100$ in Fig.~\ref{fig:dcl}.

In summary, BAO measurements of the (comoving) distance are most 
affected by $\omega_m$ and $\omega_b$, which determine 
the standard ruler of the sound horizon at the last scattering surface, 
and by $\Omega_k^\prime$, which relates the underlying 
comoving distance to the angular diameter distance. WL
distance measurements are most affected by $\omega_m$ and $\Delta_R^2$,
which effectively determine the amplitude of the lensing potential
power spectrum, and by $n_s$, whereas WL measurements of the growth 
factor do not require strong priors on cosmological parameters.


\end{document}